\documentclass[twocolumn,twocolappendix]{aastex631}
\usepackage[utf8]{inputenc}
\usepackage{enumitem}
\usepackage{graphicx}
\usepackage{hyperref}
\usepackage{mathtools}
\usepackage{comment}
\usepackage{amsmath}
\usepackage{ amssymb }
\usepackage{lipsum} 
\usepackage{float}
\usepackage{commath}
\usepackage{longtable}
\usepackage{cleveref}
\usepackage[caption=false]{subfig}
\usepackage{placeins}
\usepackage[ruled,lined]{algorithm2e}

\usepackage{xcolor}

\begin{document}
\title{Charting Galactic Accelerations with Stellar Streams and Machine Learning} 
\begin{abstract}
    We present a data-driven method for reconstructing the galactic acceleration field from phase-space (position and velocity) measurements of stellar streams. Our approach is based on a flexible and differentiable fit to the stream in phase-space, enabling a direct estimate of the acceleration vector along the stream. Reconstruction of the local acceleration field can be applied independently to each of several streams, allowing us to sample the acceleration field due to the underlying galactic potential across a range of scales. Our approach is methodologically  different from previous works, since a model for the gravitational potential does not need to be adopted beforehand. Instead, our flexible neural-network-based model treats the stream as a collection of orbits with a locally similar mixture of energies, rather than assuming that the stream delineates a single stellar orbit. Accordingly, our approach allows for distinct regions of the stream to have different mean energies, as is the case for real stellar streams. Once the acceleration vector is sampled along the stream, standard analytic models for the galactic potential can then be rapidly constrained. We find our method recovers the correct parameters for a ground-truth triaxial logarithmic halo potential when applied to simulated stellar streams. Alternatively, we demonstrate that a flexible potential can be constrained with a neural network, and standard multipole expansions can also be constrained. Our approach is applicable to simple and complicated gravitational potentials alike, and enables potential reconstruction from a fully data-driven standpoint using measurements of slowly phase-mixing tidal debris. 
    \vspace{1cm}
\end{abstract}

\author[0000-0001-8042-5794]{Jacob Nibauer}\email{jnibauer@princeton.edu}
\affiliation{Department of Astrophysical Sciences, Princeton University, Princeton, NJ 08544, USA}

\author{Vasily Belokurov}
\affiliation{Institute of Astronomy, Madingley Rd, Cambridge, CB3 0HA, UK}
\affiliation{Center for Computational Astrophysics, Flatiron Institute, 162 5th Avenue, New York, NY 10010, USA}

\author{Miles Cranmer}
\affiliation{Department of Astrophysical Sciences, Princeton University, Princeton, NJ 08544, USA}

\author{Jeremy Goodman}
\affiliation{Department of Astrophysical Sciences, Princeton University, Princeton, NJ 08544, USA}

\author{Shirley Ho}
\affiliation{Center for Computational Astrophysics, Flatiron Institute, 162 5th Avenue, New York, NY 10010, USA}
\affiliation{New York University, New York, NY 10010, USA}
\affiliation{Department of Astrophysical Sciences, Princeton University, Princeton, NJ 08544, USA}
\affiliation{Carnegie Mellon University, Pittsburgh, PA 15289, USA}

\section{Introduction}\label{sec:intro}
Stellar streams are the remnants of tidal disruption, a phenomenon which occurs when an ensemble of stars becomes tidally stripped in the underlying galactic field. Stream progenitors range from satellite dwarf galaxies (e.g. \citealt{Majewski2003,2021ApJ...920...51M,2021ApJ...920...10P}) to globular clusters (e.g. \citealt{Odenkirchen2001,2020ApJ...898L..37Y,2020MNRAS.491.5693A}). During disruption, stars lost from the progenitor follow galactocentric orbits similar to that of the progenitor itself. This results in long stream-like structures extending many degrees on the sky (see, e.g., \citealt{2018ApJ...862..114S}), sometimes wrapping around the parent galaxy multiple times \citep{Koposov2012,Belokurov2014,2020A&A...638A.104R}.

Kinematically cold, metal-poor streams are particularly useful for studies of the galactic gravitational potential, as they provide a snapshot of an extended stellar orbit in phase-space. There are several independent methods for  constraining the galactic potential: fitting orbits directly to stream measurements (e.g. \citealt{1999ApJ...512L.109J, 2006MNRAS.366.1012F, 2010ApJ...712..260K, 2011MNRAS.417..198V}), particle ejection methods \citep{2012MNRAS.420.2700K, 2015MNRAS.452..301F, 2014ApJ...795...94B,2014MNRAS.445.3788G}, action-angle tracks \citep{Sanders2014model, 2014ApJ...795...95B}, clustering in integrals-of-motion space \citep{2013MNRAS.433.1826S, 2015ApJ...801...98S, 2022MNRAS.509.5365R}, or, finally, direct $N$-body simulations \citep[e.g.][]{Dehnen2004,Law_2010}. While $N$-body simulations provide the most accurate kinematic depiction of stellar streams, they are too costly to search the parameter space efficiently for even a semi-realistic galactic potential model. Alternatively, orbit fitting of streams is attractive in its relative computational efficiency, but while the track of a stream system is similar to the orbit of the progenitor, stellar streams do not generally delineate a perfect orbit \citep{2013MNRAS.433.1813S}. Particle-release methods attempt to model the formation of stream systems by releasing particles in a trial orbit according to the (analytic) tidal radius of the progenitor system \citep{2012MNRAS.420.2700K, 2014MNRAS.445.3788G, 2015MNRAS.452..301F, Bowden2015}. After a given integration time, the final generated stream can be compared to the observed system. This method is far less costly than direct $N$-body simulations, though it still scales poorly to complicated potentials that require many model parameters. Action-angle coordinates provide a natural basis for stream formation, and can be used to obtain the model parameters of a given potential that 
 maximize clustering in action space \citep{2013MNRAS.433.1826S,2015ApJ...801...98S,2022MNRAS.509.5365R}. Related methods include modeling the action-angle distribution of streams directly. For instance, \cite{Sanders2014model} constructs a generative model for stream formation in action-angle space. Conversion from action-angle space to phase (i.e., position-velocity) space relies upon the choice of the potential. Similarly, \cite{2014ApJ...795...95B} models the initial action-angle distribution of tidal debris, which can easily be integrated forward. In this framework, the orbit of the progenitor is estimated and the mean difference in orbital frequencies along the stream relative the progenitor is characterized. The mean stream track in action-angle space is converted to phase-space using an approximate linear interpolation method. Both of these methods require a prescription for stream formation in action-angle space, and assume a uniform distribution in stripping times for stars that become unbound from the progenitor. In general, analytic transformations from phase-space coordinates $(\boldsymbol{x},\boldsymbol{v})$ to action-angle coordinates $(\boldsymbol{\theta}, \boldsymbol{J})$ are known for only a limited number of potentials for which the Hamilton-Jacobi equation is separable. Action-angle methods therefore require relatively simple models for the underlying potential, the most flexible of which is the two-component Stäckel model \citep{stackel1891integration}, which assumes that all orbits are defined by the same foci. This constraint has been shown to be incorrect for real galaxies \citep{1989MNRAS.239..571K, 2012MNRAS.426.1324B}. The Stäckel-Fudge  method \citep{2012MNRAS.426.1324B, 2015MNRAS.447.2479S} enables greater flexibility, by treating the true potential as locally similar to a Stäckel potential for which action-angle coordinates can be derived. This approach is typically implemented in the previously-mentioned works that model the action-angle distribution of tidal streams. While useful, the Stäckel-Fudge approach is approximate and will suffer if the local potential is poorly described by a Stäckel model. 

 \begin{figure*}[htp!]
    \centering
    \includegraphics[scale=0.7]{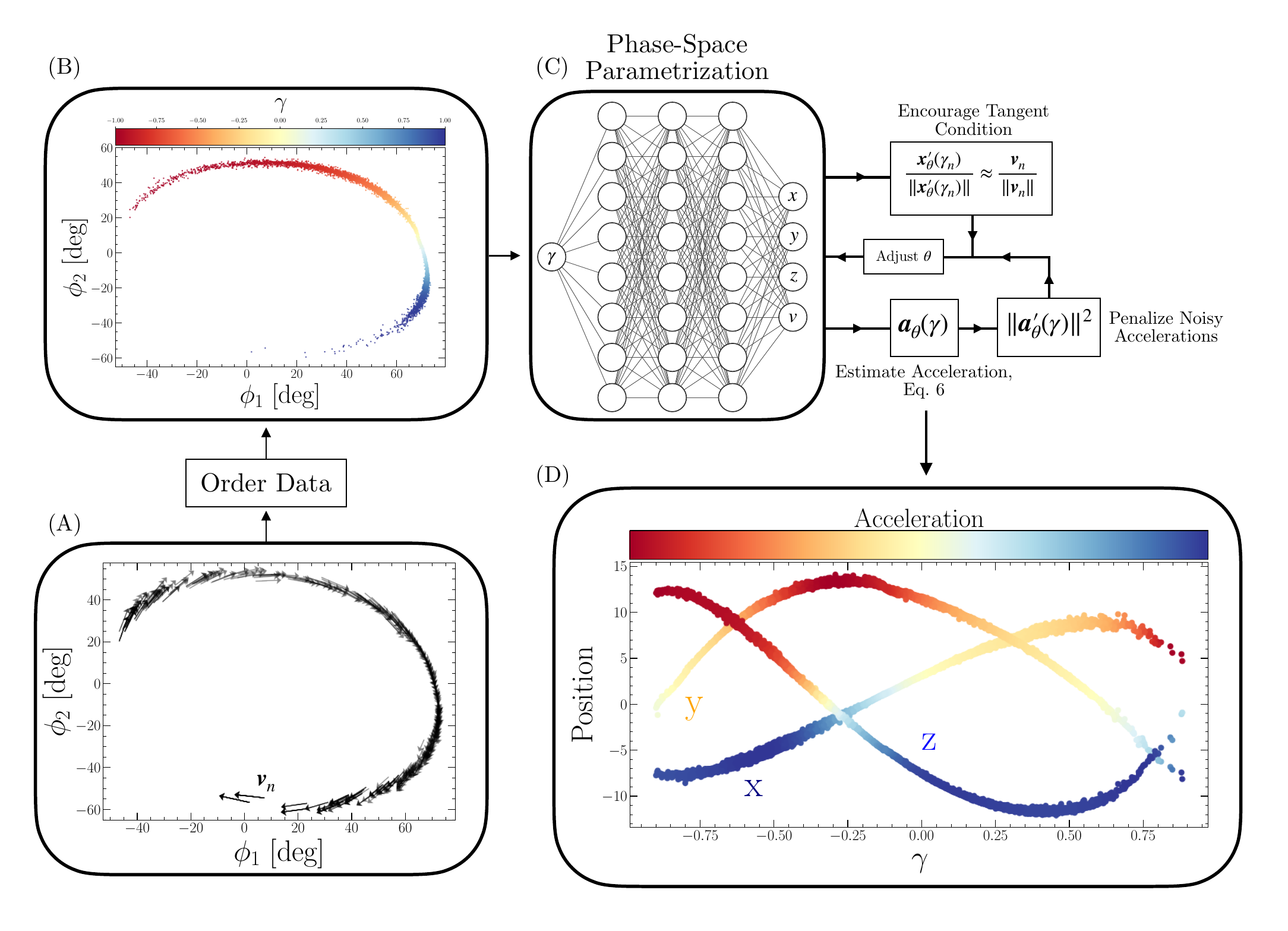}
    \caption{Illustration of our method and the underlying workflow. We plot an on-sky projection of a mock stellar stream in the bottom left panel (A). The arrows in this panel indicate the projected velocity vectors of stars along the stream, where the $n^{\rm th}$ star has velocity $\boldsymbol{v}_n$. An ordering is then assigned to stars that populate the stream, as illustrated in the top left panel (B). Stars are color-coded by the phase angle $\gamma$, which increases monotonically along the stream from the trailing to the leading tail. The positions and velocities of stars along the stream are then fed to a neural network (C), which parametrizes a mean track, $\boldsymbol{x}_\theta(\gamma)$, through the position of stream stars along with a track speed. Both the stream track and track speed are parametrized in terms of the $\gamma$-parameter. These fits can be differentiated and combined through Eq.~\ref{eq: acceleration_eqn} to estimate the cartesian components of the acceleration vector along the stream directly. An intermediate model for the galactic potential never needs to be specified. During training of the neural network, we encourage the condition that the tangent vector to the stream track falls roughly parallel to the local trajectory of stream stars. We also encourage a smooth acceleration field, by penalizing neural-network parameters that give rise to large local changes in accelerations along the stream track. An example output is illustrated in the bottom right panel (D), where each cartesian component of the stellar stream is ``unwrapped" along the $\gamma$-axis, and color-coded by the inferred accelerations.}
    \label{fig: workflow}
\end{figure*} 

 Regardless of the method and trade-offs therein, most studies using stellar streams to constrain the galactic potential rely on a narrow class of analytic models or truncated basis function expansions to represent the potential. However, the true morphology of the dark matter halo in the Galaxy could deviate from simplified models (see, e.g., \citealt{prada2019dark,2021ApJ...919..109G,Shao2021}). Adopting an incorrect model for the galactic potential can lead to biased parameter constraints, regardless of the accuracy of the adopted stream-modeling approach \citep{2014ApJ...795...94B}. A method which does not place functional priors on the underlying galactic potential could therefore provide an important advancement in mapping the detailed distribution of dark matter in e.g. the Milky Way stellar halo. Such constraints could then be compared to cosmological simulations, for which functional priors on the dark matter distribution are not adopted (e.g. \citealt{2008Natur.454..735D}). 
 
Recent work \citep{2018ApJ...867..101B} has demonstrated that more flexible potential models (e.g. basis function expansions) fit to individual stellar streams probe the local properties of the potential in the neighborhood of the given stream, but are not necessarily representative of the global potential. Consequently, a population of streams constrains different aspects of the global potential. \cite{2018ApJ...867..101B} therefore highlight the need for a flexible method that can constrain the global potential from multiple streams simultaneously, while accurately reproducing the local features of the potential probed by individual systems. Motivated by a Fisher-information analysis of simulated stellar streams, these authors suggest to analyze each stream independently using a flexible model, thereby maximizing the learned information content of the given stream. As a post-processing step, a global potential can be constrained by interpolating the individualized stream fits across multiple systems.

In this work we consider a fully flexible and differentiable approach to directly estimate the galactic acceleration field in the neighborhood of a given stream. Our method does not require a model for the underlying galactic potential, nor does it require streams to delineate isoenergy curves in phase-space (i.e. stellar orbits). The output of our analysis is similar to that of \cite{2022MNRAS.511.1609N}, since we constrain galactic accelerations rather than a latent potential model. We assume that the potential is mostly static, or changing  slowly. Once the acceleration field in the neighborhood of a number of streams is measured, we demonstrate that a global, flexible gravitational potential can be constrained. The resulting potential is consistent with the local properties of the acceleration field where each stream resides, and is fully data-driven. Alternatively, interpretable analytic models for the potential can be fit to the estimated accelerations along each stream to recover physical parameters. 

The paper is organized as follows. In \S\ref{sec: method} we introduce our method for estimating accelerations from stellar streams. We provide a probabilistic framework to connect our model to the data in \S\ref{sec: fitting_streams}, and discuss regularization of the inferred acceleration field in \S\ref{sec: physics_informed}. In \S\ref{sec: mock_streams}, we apply our method to simulated stellar stream data to demonstrate an accurate reconstruction of the acceleration field in a known ground truth potential. We compare our approach to fitting stellar streams with orbits in \S\ref{sec: orbit_fitting}, and demonstrate accurate potential reconstruction in \S\ref{sec: potential}. We provide a discussion of our method in \S\ref{sec: discussion}, along with limitations of this work and future directions.

\section{Method}\label{sec: method}
We now introduce our approach for estimating the galactic acceleration field from 6D phase-space measurements of stellar streams. Several streams in the Milky Way have 6D phase-space measurements available (e.g., \citealt{2010ApJ...712..260K, 2020A&A...635L...3A}), though the majority have only partial coverage. As a first analysis, we utilize the full 6D phase-space distribution of stellar streams. We will consider the case of missing phase-space observations in a future work. We provide a brief summary of our approach below, and a more technical description in \S\ref{sec: technical_description}.

Stellar streams are populated by stars on proximate but distinct orbits. Diversity among the orbits depends upon the potential of the progenitor, its history of mass loss, the mean potential of the galactic host, and perhaps on substructure within the host’s halo. Absent the last of these, one expects the local mean of the orbits to vary slowly along each arm of the stream.  Motivated by these considerations, in this work we treat stellar streams as a mixture of orbits, with neighboring segments of the stream populated by stars with a similar mixture. However, distinct segments of the stream may have different mixtures of orbits. Accordingly, we do not assume that streams delineate a single orbit in the correct galactic potential. Instead, we parametrize a path through a given stream in phase-space which characterizes the average local trajectory of the stellar orbital mixture. This path is flexible and differentiable, and does not necessarily coincide with any single orbit in the galactic potential. By applying the chain rule to eliminate explicit time-dependence, an acceleration vector can be estimated at each evaluation point along the stream from the parametrized path. An intermediate model for the galactic potential never needs to be specified. 

Our analysis workflow is illustrated conceptually in Fig.~\ref{fig: workflow}. In the bottom left panel (A), we plot a projection of a mock stream in angular on-sky coordinates $(\phi_1,\phi_2)$. Arrows indicate velocity vectors. The track of the stream is determined by linking nearest neighbors in 6D phase-space. This induces an ordering along the stream, which we encode with a scalar parameter $\gamma \in [-1,1]$ (B). This enables us to ``unwrap" a stream along the $\gamma$-axis, similar to unwrapping a particle's trajectory, $(x(t), \ y(t))$, in time. In practice, the parameter can be determined using an on-sky position angle and velocities to distinguish the leading arm from the trailing arm. We delay a more general discussion of the $\gamma$-parameter and its determination to
\S\ref{sec: gamma}.

Once the data are ordered, a neural network is used to parametrize a flexible and differentiable track along the observed stream (Fig.~\ref{fig: workflow}; panel C). This neural network takes $\gamma$ as an input, and outputs the three cartesian coordinates $(x,y,z)$ and speed $v$ of the track. We denote the position-space stream track with  $\boldsymbol{x}_{\theta}(\gamma)$, where $\theta$ are the parameters of the neural network. The mean speed as a function of $\gamma$ is $v_{\varphi}(\gamma)$, where $\varphi$ are the parameters of an independent neural network that characterizes the track speed. In Fig.~\ref{fig: workflow} we consolidate these neural networks down to one system, though the usage of two independent neural networks for the stream track and track speed is only an implementation detail. 

We assume that locally, stream stars are characterized by a similar mixture of orbits at a slightly different orbital phase. In order to maximize the validity of this assumption, neural-network training is carried out so that the stream track is encouraged to point along the local mean motion of stars along the stream. This condition helps the neural network determine a path through the stream for which our assumptions are most completely satisfied. In addition,  we adopt a prior on the neural-network parameters $\theta$ that encourages smooth estimates of the acceleration field as a function of the scalar parameter $\gamma$. This prior is implemented by penalizing neural-network parameters that give rise to large derivatives of estimated accelerations with respect to $\gamma$. Parameters of the stream-track neural network are adjusted so as to satisfy the tangent and smoothness conditions while also remaining compatible with the observed phase-space data. 
 
 An example output of our analysis is shown in the bottom right panel of Fig.~\ref{fig: workflow} (D), where stream stars are ``unwrapped" along the $\gamma$-axis and color-coded by their inferred accelerations in each cartesian component. In summary, our analysis utilizes a neural network to model the track of a stream in $6D$ phase-space. By applying the chain rule, derivatives of position and velocity along this track provide direct estimates of galactic accelerations. The required inputs are the positions and velocities of stream-members.  

\subsection{Technical Description}\label{sec: technical_description}
\begin{figure*}[htp!]
    \centering
    \includegraphics[scale=0.55]{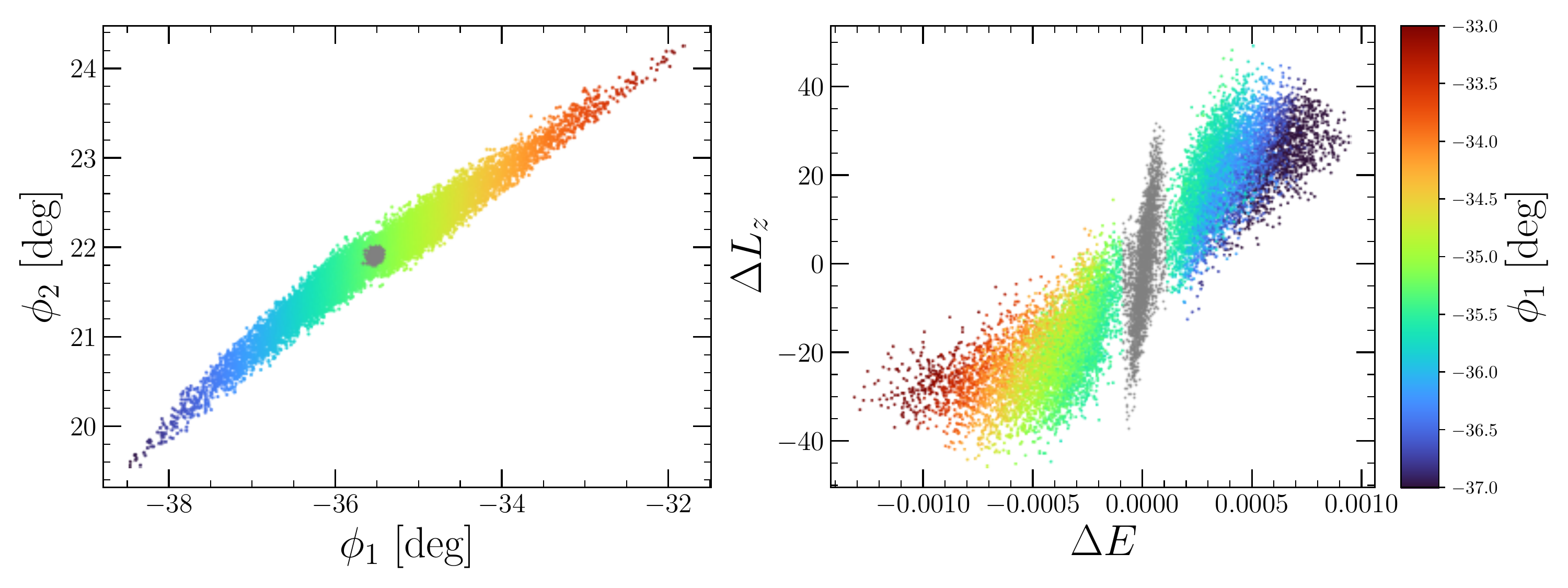}
    \caption{$N$-body realization of a stellar stream in an axisymmetric halo potential. In the left panel, we plot a projection of the $N$-body stream in angular coordinates. In the right panel, the stream is plotted in the space of energy and the z-component of angular momentum, both of which are integrals of motion in a stationary axisymmetric potential. These quantities are plotted relative to the progenitor cluster, shaded in gray. Outside of the progenitor, tracers are color-coded by their position angle, $\phi_1$, in the left panel. This angle increases monotonically along the stream, demonstrating that stream stars have locally similar integrals of motion at slightly different phase angles. This figure is provided only for conceptual purposes; our method does not model the energy distribution of stellar streams directly, nor does it work in the space of integrals of motion.}
    \label{fig: Stream_Sample}
\end{figure*}

Any given stream consists of a collection of stars on somewhat different orbits. In the space of Integrals of Motion (IoM), stream tracers are clustered with locally similar IoM at slightly different phase angles (see, e.g., \citealt{2013MNRAS.433.1826S,2015MNRAS.452..301F,2021MNRAS.502.4170R}). However, at a global level stream tracers span a range of energies. We demonstrate this property in Fig.~\ref{fig: Stream_Sample}, where we plot an $N$-body realization of a stellar stream generated in an axisymmetric potential after $\sim 3.5~\rm{Gyr}$ of evolution (left panel). For an axisymmetric potential, the energy ($E$) and z-component of angular momentum ($L_z$) are IoM. We plot the $N$-body stream in the space of binned mean energy and binned mean $L_z$ in the right panel of Fig.~\ref{fig: Stream_Sample}, where these quantities are relative to the progenitor cluster. Points are color-coded by the average angular coordinate $\phi_1$ in the left panel, which increases monotonically along the stream. The progenitor cluster is shaded in gray. In the space of IoM, we see a distinctive ``bow-tie" feature, with separate wings corresponding to the leading and trailing arms of the tidal stream. The ``bow-tie" is a result of the progenitor oscillating between pericenter and apocenter while losing stars to the galactic tidal field. Once stars become unbound from the progentior, they are ``locked-in" to the potential of the galaxy and no longer oscillate with the progenitor. Further discussion is provided in \citet{2014MNRAS.445.3788G} and \citet{ 2011ApJ...731...58Y}.  

From this figure, it is clear that tracers are---on average---sorted by their energy and angular momentum along the stream. That is, stream stars farthest from the progenitor typically have the largest offsets in $E$ and $L_z$ from the progenitor, whereas unbound stars near the progenitor have smaller offsets in these quantities \citep[see, e.g.,][]{Johnston2001,2014MNRAS.445.3788G}.

For a given band in $\phi_1$, Fig.~\ref{fig: Stream_Sample} illustrates that the stream is populated by stars each with slightly different orbits. However, the local mixture of orbits changes only gradually along $\phi_1$ from one segment of the stream to the next. Consequently, while stellar streams do not trace a single orbit in the galactic potential, they typically consist of an ensemble of stars characterized by a locally similar mixture of orbits. At a global level, however, any two segments of the stream separated by a substantial phase angle can have larger differences in orbits. 

Our analysis exploits this property of stellar streams: that is, we assume adjacent small segments of the stream are populated by an ensemble of stars with a  similar distribution in the space of IoM. The properties of this distribution are allowed to evolve along the stream, albeit slowly. If two local segments of the stream have a similar distribution in the space of IoM, then the mean of the IoM distribution will not vary significantly with small changes in phase angle along the stream. Under this view, the mean path of a stream is characteristic of many small orbital segments which have locally similar IoM. 

The local clustering of IoM in Fig.~\ref{fig: Stream_Sample} (bottom right) as a function of position angle $\phi_1$ motivates us to fit the mean dynamical properties of the stream as a function of some underlying phase. We refer to this fit as the stream track, which characterizes the local position and motion of stream tracers as a function of e.g. position angle $\phi_1$. The stream track forms a curve in 6-dimensional phase-space. We parametrize this curve in terms of the scalar parameter $\gamma$:
\begin{equation}\label{eq: x_v}
    \begin{split}
        \boldsymbol{x} &= \boldsymbol{x}(\gamma) \\
        \boldsymbol{v} &= \boldsymbol{v}(\gamma),
    \end{split}
\end{equation}
where $\boldsymbol{x}$ and $\boldsymbol{v}$ are position and velocity vectors, respectively. In practice, we estimate $\boldsymbol{x}(\gamma)$ and $\boldsymbol{v}(\gamma)$ using a neural network with parameters $\theta$. We discuss estimating these quantities from phase-space data of stellar streams in \S\ref{sec: fitting_streams}. However, in this section we suppress dependence on $\theta$ for simplicity. Typically, $\gamma$ is taken to be a position angle on the sky, increasing from the trailing to leading arm of the stream (e.g., $\phi_1$ in \citealt{2010ApJ...712..260K} and  Fig.~\ref{fig: Stream_Sample}, or $\Lambda$ in  \citealt{2021MNRAS.501.2279V}). 

Our central assumption is that stellar streams have locally similar mixtures of IoM at slightly different phase angles. That is, the mean instantaneous motion of a small stream segment will fall roughly parallel to the local direction of the stream track. Mathematically, this assumption implies the existence of a relationship between the parametrized stream track in position space, $\boldsymbol{x}(\gamma)$, and the parametrized track velocity, $\boldsymbol{v}(\gamma)$. In particular, we define the unit vector tangent to the stream track at scalar parameter $\gamma$ as
\begin{equation}
    \boldsymbol{{T}}(\gamma) \equiv
    \frac{\boldsymbol{x}^{\prime}(\gamma)}{\Vert \boldsymbol{x}^{\prime}(\gamma)\Vert},
\end{equation}
where $\boldsymbol{x}^{\prime}(\gamma) \equiv d\boldsymbol{x}(\gamma)/d\gamma$. If the velocity of an infinitesimal segment of the stream track follows the local morphology of the stream, this means
\begin{equation}\label{eq: velo}
    \boldsymbol{v}(\gamma) = \boldsymbol{T}(\gamma) \Vert \boldsymbol{v}(\gamma) \Vert,
\end{equation}
where $\Vert \boldsymbol{v}(\gamma) \Vert$ is the local mean speed of stream tracers at scalar parameter $\gamma$. In practice, this relationship (Eq.~\ref{eq: velo}) between $\boldsymbol{x}(\gamma)$ and $\boldsymbol{v}(\gamma)$ is enforced through our modeling, such that the best fit stream track is the one for which our assumptions are most completely satisfied given the observed data (see \S\ref{sec: fitting_streams} for details).

We emphasize that while Eq.~\ref{eq: velo} is valid for a stellar orbit in the galactic potential, it does not necessarily imply that the parametrized track $(\boldsymbol{x}(\gamma), \boldsymbol{v}(\gamma))$ is that of {\it any} particular orbit in the underlying potential. In fact, provided that the mean IoM of neighboring small patches of a stream are locally similar, the parametrized track as a whole is not required to delineate a path of constant energy in {\it any potential} whatsoever. This is because while stream tracers typically fall along similar orbits locally, distinct segments of a stream separated by substantial phase angles may have significantly different energies (as well as other IoMs) and therefore different orbits. We demonstrate this major advantage of our analysis in \S\ref{sec: orbit_fitting}. 

We next define the scalar path length $s$, which represents the path of the stream track with dimensions of length. Explicitly, this quantity can be expressed as a function of the scalar parameter $\gamma$ as follows:
\begin{equation}\begin{split}\label{eq: path_length_gamma}
    s(\gamma) &= \int\limits_{0}^{\gamma} d\tilde{\gamma} \sqrt{\left(\frac{dx}{d\gamma} \Big\vert_{\tilde{\gamma}}\right)^2 + \left(\frac{dy}{d\gamma}\Big\vert_{\tilde{\gamma}} \right)^2 + \left(\frac{dz}{d\gamma}\Big\vert_{\tilde{\gamma}}\right)^2   } \\
    &= \int\limits_{0}^{\gamma} d\tilde{\gamma} \sqrt{\frac{d\boldsymbol{x}}{d\gamma}\Big\vert_{\tilde{\gamma}} \cdot \frac{d\boldsymbol{x}}{d\gamma}\Big\vert_{\tilde{\gamma}}},
    \end{split}
\end{equation}
where we have used $\boldsymbol{x} = (x,y,z)$ and $\tilde{\gamma}$ is the integration variable.
Following the discussion above, we identify $\boldsymbol{x}(\gamma)$ locally with an average particle trajectory $\boldsymbol{x}(t)$, the velocity $\boldsymbol{v}(\gamma)$ with $d\boldsymbol{x}/dt$, and $ds/dt$ with $\Vert \boldsymbol{v}(t)\Vert$.
Applying the chain rule, the local correspondence between $\gamma$ and time is then
\begin{equation}
    \frac{d\gamma}{dt}\equiv \frac{\Vert\boldsymbol{v}\Vert}{ds/d\gamma}\,.
\end{equation}
With this correspondence, the local particle acceleration is obtained as\begin{linenomath}
\begin{equation*}
    \boldsymbol{a}\equiv\frac{d\boldsymbol{v}}{dt}=\frac{d\boldsymbol{v}}{d\gamma}\frac{d\gamma}{dt}\to \frac{d\boldsymbol{v}}{d\gamma}\frac{\Vert\boldsymbol{v}\Vert}{ds/d\gamma}.
\end{equation*}\end{linenomath}
In other words,
\begin{equation}\label{eq: acceleration_eqn}
    \boldsymbol{a}(\tilde{\gamma}) = \frac{d\boldsymbol{v}}{d\gamma}\Big\vert_{\tilde{\gamma}} \left(\frac{d\boldsymbol{x}}{d\gamma}\Big\vert_{{\tilde{\gamma}}} \cdot \frac{d\boldsymbol{x}}{d\gamma}\Big\vert_{{\tilde{\gamma}}} \right)^{-1/2} \Vert{}\boldsymbol{v}(\tilde{\gamma}) \Vert{},
\end{equation}
where $\tilde{\gamma}$ is an evaluation point, and the term in parentheses is from Eq.~\ref{eq: path_length_gamma}. Consequently, given a differentiable parametrization of the stream track in phase-space, the acceleration vector along the track can be estimated using Eq.~\ref{eq: acceleration_eqn} without having to formulate a model for the potential. This enables a substantial increase in flexibility over previous methods, which require an analytic form for the potential to be adopted. In this way, our approach is agnostic to the functional form of the potential. Furthermore, Eq.~\ref{eq: acceleration_eqn} is inherently local and does not assume that distinct segments of a stream separated by a substantial phase fall along the same orbit. We demonstrate this quality in subsequent sections. 

With Eq.~\ref{eq: acceleration_eqn}, stellar streams can be used to sample the galactic acceleration field directly. Samples of the galactic acceleration field can be utilized to infer the potential, $\Phi$, through the relation
\begin{equation}\label{eq: a_nabla_phi}
    \boldsymbol{a}(\gamma) = -\nabla_{\boldsymbol{x}}{\Phi} \Big\vert_{\boldsymbol{x}(\gamma)}.
\end{equation}

\subsection{Fitting Stellar Streams}\label{sec: fitting_streams}
From a set of 6D phase-space observations (that is, positions and velocities of stars belonging to the stream) a number of analytical and data-driven methods can be employed to estimate a differentiable parametrized curve $(\boldsymbol{x}(\gamma), \boldsymbol{v}(\gamma))$ in phase-space. These range from polynomial fits, splines, to fully flexible neural network representations. In the subsequent sections we consider the latter of these methods, since a neural network representation allows us to flexibly describe the data while encouraging smooth estimates of the underlying acceleration field. In this section, we also assume that a suitable $\gamma$ for each tracer has been assigned, though we discuss a more general determination of the $\gamma$  parameter in \S\ref{sec: gamma}. The set of measurements required for the present analysis is then $\mathcal{D} =\{(\boldsymbol{x}_n, \boldsymbol{v}_n, \gamma_n)\}_{n=1}^N$, where $\boldsymbol{x}_n$ and $\boldsymbol{v}_n$ denote the position and velocity of the $n^{\rm th}$ stream-member, and $\gamma_n$ encodes where the stream-member belongs along the unwrapped stream. There are $N$ total stream members.

Given the separability of the tangent vector $\boldsymbol{T}$ from the stream track speed $\Vert \boldsymbol{v}\Vert$ in Eq.~\ref{eq: velo}, we fit the stream track $\boldsymbol{x}_{\theta}(\gamma)$ and speed ${v}_{\varphi}(\gamma)$ separately, where $\theta$ and $\varphi$ are the parameters of the two independent neural network representations of these quantities. We use a standard class of neural networks called multilayer perceptron (MLP) with a fully connected feed-forward architecture. The neural network $\boldsymbol{x}_{\theta}$ maps a scalar input, $\gamma$, to a 3-dimensional position output $(x,y,z)$. The network $v_{\varphi}(\gamma)$---which we refer to as the speed neural network, or track speed---maps the same scalar input $\gamma$ to a scalar output, representing the local speed of stream tracers. We discuss the specific architecture of these neural networks in Appendix~\ref{App: Architectures}. 

Fitting a MLP involves tuning the parameters of the neural network until an objective function is optimized, typically analogous to $\chi^2$ fitting. For the case of the speed neural network, $v_{\varphi}(\gamma)$, we adopt a mean-squared-error loss function of the form
\begin{equation}\label{eq: speed_loss}
    \ell_{\varphi}\left(\varphi\right) = \frac{1}{N}\sum_{n=1}^N w_n \left( {v_{\varphi}}(\gamma_n) - {v_n} \right)^2 + h \Vert \varphi \Vert^2_2,
\end{equation}
where $w_n$ is a weight for each measurement that could be informed by estimated signal-to-noise of a given data point. The term $\Vert \varphi \Vert^2_2$ denotes the $L_2$ norm of the neural network parameters $\varphi$, and $h$ is a hyper-parameter of the model. Practically, the $L_2$ norm works to enforce smoothness in the space of $v_{\varphi}$, and reduce overfitting by penalizing particularly large parameter values. In the machine-learning literature, this regularization technique is commonly referred to as weight decay. We adopt $h\sim 10^{-6}$, where $h$ is the weight decay parameter. The optimal parameters $\hat{\varphi}$ are determined using standard back-propagation routines with the \texttt{Adam} optimizer \citep{2014arXiv1412.6980K}, until the loss function in Eq.~\ref{eq: speed_loss} is minimized. All neural-network training in this analysis is performed in mini-batches, such that model parameters are only updated after iterating through several observations (see \citealt{2018arXiv180407612M} for a review on mini-batch training). This is standard practice when training neural networks, and reduces computational load while increasing the efficiency of training. 

We next turn to our determination of the stream track, $\boldsymbol{x}_{\theta}(\gamma)$. Eq.~\ref{eq: velo} implies that the stream track is parallel to the mean velocity of the local stream stars. This follows from our assumption that a given segment of the stream has a similar mixture of IoM compared to an adjacent segment. We build this property into our modeling, in order to ensure that the best-fit stream track is the one that most completely satisfies our assumptions. We first define the quantity
\begin{equation}\label{eq: track_tangent}
    \boldsymbol{T}_{\theta}(\gamma) \equiv
    \frac{\boldsymbol{x}_{\theta}^{\prime}(\gamma)}{\Vert \boldsymbol{x}_{\theta}^{\prime}(\gamma)\Vert},
\end{equation}
which represents the unit-vector tangent to the stream track at scalar parameter $\gamma$. We subscript this quantity with $\theta$, since it depends on the parameters of the stream track neural network. The observed unit velocity vector for the $n^{\rm th}$ stream star is $\boldsymbol{T}_n \equiv \boldsymbol{v}_n / \Vert \boldsymbol{v}_n \Vert$.

\begin{linenomath}
Generally, the objective function can be thought of as a likelihood for the data $\mathcal{D}$ given a set of model parameters (e.g., $\theta$). Indeed, Eq.~\ref{eq: speed_loss} can be equivalently expressed as the combined negative log-likelihood of Gaussian distributed random variables, though presenting the objective function as a weighted mean-squared-error loss as we have done is more typical in the machine-learning literature. For the case of the stream track neural network $\boldsymbol{x}_{\theta}$, we obtain an objective function using the more general Bayesian framework of likelihoods and priors, since we later incorporate a regularization prior in our analysis. We connect the data $\mathcal{D}$ to the model parameters $\theta$ through a multivariate Gaussian distribution. The likelihood for the $n^{\rm th}$ stream star is 
\begin{multline}\label{eq: single_star_likelihood}
    P\left( \begin{bmatrix} 
    \boldsymbol{x}_n \\
    \boldsymbol{T}_n
    \end{bmatrix} \Bigg\vert \theta, \tau_1, \tau_2
    \right) = \\ \mathcal{N}\left(\begin{bmatrix} 
    \boldsymbol{x}_n \\
    \boldsymbol{T}_n
    \end{bmatrix} \Bigg\vert \begin{bmatrix} 
    \boldsymbol{x}_\theta(\gamma_n) \\
    \boldsymbol{T}_\theta(\gamma_n)
    \end{bmatrix}, \begin{bmatrix} 
    \tau_1 \boldsymbol{I}_{3\times3} & \boldsymbol{0} \\
    \boldsymbol{0} & \tau_2 \boldsymbol{I}_{3\times3} 
    \end{bmatrix}
    \right),
\end{multline}
\end{linenomath}
where $\mathcal{N}(\boldsymbol{x} \vert \boldsymbol{\mu}, \boldsymbol{C})$ is a multivariate Gaussian distribution with mean vector $\boldsymbol{\mu}$ and covariance matrix $\boldsymbol{C}$, evaluated at $\boldsymbol{x}$. The symbols $\boldsymbol{I}_{3\times 3}$ denote $3\times3$ identity matrices. We treat $\tau_1$ and $\tau_2$ as constant hyper-parameters of the model, which act as weights on the real space position and trajectory of the parametrized track, respectively. Similar to the weights $w_n$ in Eq.~\ref{eq: speed_loss}, the likelihood in Eq.~\ref{eq: single_star_likelihood} provides a pathway to incorporate errors on the phase-space measurements used in our fits through the covariance term, analogous to the likelihood in \cite{2014ApJ...795...94B,2018arXiv180407766H}. We postpone the inclusion of errors to a future work.

Assuming that the measurement of position and velocity for each star is statistically independent, we may write the likelihood for the data as a product over all stars:
\begin{equation}\label{eq: prob_D_given_theta}
    \mathcal{L}\left(\mathcal{D} | \theta, \tau_1, \tau_2\right) = \prod_{i=1}^N P\left( \begin{bmatrix} 
    \boldsymbol{x}_n \\
    \boldsymbol{T}_n
    \end{bmatrix} \Bigg\vert \theta, \tau_1, \tau_2
    \right).
\end{equation}
The logarithm of this quantity can be taken as an objective function to be maximized over the model parameters $\theta$. We derive the explicit objective function used during model training in \S\ref{sec: physics_informed}, where we include the addition of regularization prior which encourages smooth estimates of the acceleration field. 

While fitting for both the position and trajectory of the stream track simultaneously in Eq.~\ref{eq: prob_D_given_theta} might appear redundant, it has the beneficial feature of self-regulation. That is, we fit for a parametrized curve in phase-space with the property that the tangent vector at each point along the curve is parallel to the local instantaneous motion of stream tracers in 3D space. Therefore, the best fit stream track is not required to pass through the the centroid of the stream (i.e., the central region perpendicular to the elongation of the stream) as one would find when fitting the 3D positions of stream stars alone. Instead, the best fit stream track is the one which will satisfy our assumptions most completely. We illustrate this aspect of our analysis in \S\ref{sec: mock_streams}.

\subsection{Regularization of Inferred Accelerations}\label{sec: physics_informed}
Poisson's equation, $\nabla_{\boldsymbol{x}}^2 \Phi = 4\pi G \rho$, connects the Laplacian of the gravitational potential to the underlying matter density, $\rho$. From this relation, in order for the acceleration field to be physical it must have the property $\nabla_{\boldsymbol{x}} \cdot \boldsymbol{a} \leq 0$. Furthermore, because the force of gravity is conservative we also expect $\nabla_{\boldsymbol{x}} \times \boldsymbol{a} = \boldsymbol{0}$; this ensures the existence of a scalar potential $\Phi$ such that $\boldsymbol{a} = -\nabla_{\boldsymbol{x}} \Phi$. We have experimented with incorporating similar constraints as physically motivated priors on the neural network parameters that characterize the stream track and track speed. However, the method presented in this paper works
by estimating changes in position and velocity along the 1-dimensional track of a stream, which we relate to the underlying galactic acceleration field. Our estimate of the acceleration field, while 3-dimensional, is also confined to this track. As a result, we infer what is effectively a ``slice" through the full acceleration field, parametrized by $\boldsymbol{x}_{\theta}(\gamma)$. Therefore, from our method alone and without further assumptions on the functional form of the potential, we only have access to derivatives of the acceleration field along the stream track. We cannot in general calculate $\nabla_{\boldsymbol{x}} \cdot \boldsymbol{a}$ at each evaluation point along the stream track. 
With regards to enforcing a physical mass density $\rho \geq 0$ along the stream track, in Appendix~\ref{app: physical_track} we show that this condition can always be satisfied independent of the shape, location, or velocity of the track. This is again a consequence of our analysis providing a slice through the acceleration field, rather than a full spatial map. However, estimates of the acceleration field along one or more streams can be combined to constrain a flexible potential. From this potential, one can compute $\nabla_{\boldsymbol{x}} \cdot \boldsymbol{a}$ and other expressions involving full-spatial derivatives. We demonstrate this aspect of our analysis in \S\ref{sec: potential}.

As an alternative and more feasible regularization, we adopt a prior on the neural network parameters that penalizes noisy estimates of the acceleration field. That is, we down-weight neural network parameters for which $\Vert d\boldsymbol{a}_\theta/{d\gamma}\Vert$ is large. 

The acceleration vector inferred from Eq.~\ref{eq: acceleration_eqn} depends on two independent neural networks: namely, the stream track $\boldsymbol{x}_{\theta}(\gamma)$, and the track speed $v_{\varphi}(\gamma)$. While it is possible to train both of these neural networks jointly so that $\Vert d\boldsymbol{a}_\theta/{d\gamma}\Vert$ is typically small, we have found this to be unnecessary and  computationally expensive. Instead, we fit the speed neural network $v_{\varphi}(\gamma)$ separately, and fix its model parameters to the optimal values $\hat{\varphi}$. We then train $\boldsymbol{x}_{\theta}(\gamma)$, penalizing large derivatives of the inferred accelerations along the stream with $\varphi$ set to the optimal values, $\hat{\varphi}$. We find that the stream track will sometimes induce noise artifacts in the inferred accelerations, presumably because it is a more complicated neural network which parametrizes a curve in 3-dimensional space. We find that such artifacts are uncommon for the simpler speed neural network, $v_{\varphi}$, provided that we regularize the model parameters of this network with a weight decay term (Eq.~\ref{eq: speed_loss}).

In the context of Bayesian inference, we promote smoothness of the acceleration field by adopting an improper exponential prior over the stream track parameter $\theta$, given the optimal track speed parameters $\hat{\varphi}$ as follows:
\begin{equation}
    P(\theta | \hat{\varphi}, \tau_3) \propto  \exp\Big\{-\frac{1}{\tau_3}\sum_{n=1}^N
    \Big\Vert \frac{d\boldsymbol{a}_{\theta | \hat{\varphi}}(\gamma_n)}{d\gamma} \Big\Vert^2
     \Big\},
\end{equation}
where $\tau_3 >0$ is a hyper-parameter of the model, and $\boldsymbol{a}_{\theta | \hat{\varphi}}$ is the acceleration vector along the stream inferred with $\theta$ free to vary and $\varphi$ fixed to the optimal values, $\hat{\varphi}$. Applying Bayes' theorem, the posterior over the parameters $\theta$ is proportional to the product between the likelihood and the prior,
 \begin{equation}\label{eq: theta_posterior}
     P(\theta | \mathcal{D}, \hat{\varphi}, \tau_1, \tau_2, \tau_3 ) \propto \mathcal{L}\left(\mathcal{D} | \theta, \tau_1, \tau_2 \right)P\left(\theta | \hat{\varphi}, \tau_3\right),
 \end{equation}\begin{linenomath}
 where the likelihood is from Eq.~\ref{eq: prob_D_given_theta}. We utilize Eq.~\ref{eq: theta_posterior} as an objective function over $\theta$. In particular, up to a constant, we minimize the negative logarithm of the posterior density over $\theta$,
\begin{multline}\label{eq: log_like_mod}
    \ell_{\theta}\left(\theta\right) \equiv  \\ \sum_{n=1}^N \Bigg[ \lambda_1\Vert \boldsymbol{x}_n - \boldsymbol{x}_{\theta}(\gamma_n)  \Vert^2  + \lambda_2 \Vert \boldsymbol{T}_n - \boldsymbol{T}_{\theta}(\gamma_n) \Vert^2 \\ + \lambda_3\Big\Vert \frac{d\boldsymbol{a}_{\theta | \hat{\varphi}}(\gamma_n)}{d\gamma} \Big\Vert^2\Bigg],
\end{multline}\end{linenomath}
 where we have defined $\lambda_1 \equiv \tau_1^{-1}$, $\lambda_2 \equiv \tau_2^{-1}$, and $\lambda_3 \equiv \tau_3^{-1}$. Minimization of Eq.~\ref{eq: log_like_mod} can be performed efficiently using standard back-propagation routines. In this work, we use the \texttt{Adam} optimizer implemented in PyTorch \citep{2014arXiv1412.6980K,2019arXiv191201703P}. In practice, we set $\lambda_1 = 1$ and initialize training with $\lambda_2 = 1$ and $\lambda_3 = 0$. After $\sim 50$ initial training epochs, we set $\lambda_2$ to a larger value, such that the order of magnitude of the vector norm $\lambda_2 \Vert \boldsymbol{T}_n - \boldsymbol{T}_{\theta}(\gamma_n) \Vert^2$ is at least the same as that of $\Vert \boldsymbol{x}_n - \boldsymbol{x}_{\theta}(\gamma_n)  \Vert^2$. This ensures that the first two terms in the loss function of Eq.~\ref{eq: log_like_mod} receive similar weight. After an additional $\sim 50$ training epochs, we then choose $\lambda_3$ such that $\lambda_3\Vert d\boldsymbol{a}_{\theta | \hat{\varphi}}/d\gamma \Vert^2$ has roughly the same order of magnitude compared to the first penalty term on average. 
 
\begin{figure*}[htp!]
    \centering
    \includegraphics[scale=0.7]{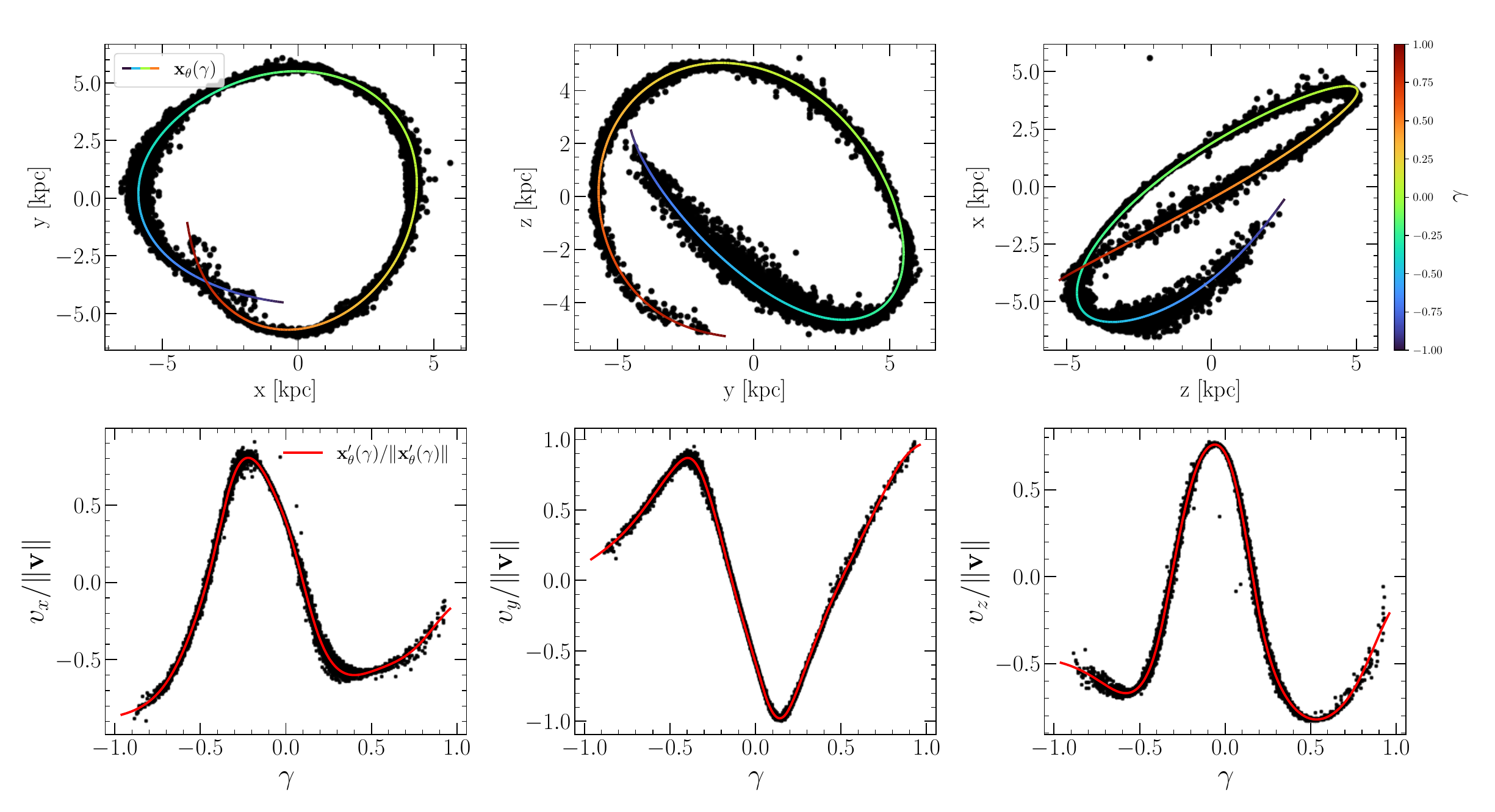}
    \caption{{\it Top Row:} Mock stellar stream plotted in galactocentric $x$, $y$, and $z$ coordinates (black scatter points). The bulk motion and position of the stream is parametrized with a flexible neural network, $\boldsymbol{x}_{\theta}(\gamma)$. We illustrate this parametrization with the colored curve, color-coded by the $\gamma$ value along the stream. {\it Bottom Row:} For the same mock stream, we plot the unit-vector trajectories of tracers in the $x$, $y$, and $z$ direction (black scatter points). The red curve is our parametrization of these points, taken as the derivative of the colorful curve in the top panel with respect to $\gamma$, and normalized. The parametrized curves in this figure are generated from a single neural network. Streams are unwrapped using the method described in Appendix~\ref{App: Unwrap}.}
    \label{fig: Stream_Parametrized}
\end{figure*}

 Our analysis relies heavily on the tangent vector to the stream track, $\boldsymbol{T}_\theta$. This can be seen from our expression for the acceleration vector along the stream track, Eq.~\ref{eq: acceleration_eqn}, and our expression for the track velocity, Eq.~\ref{eq: velo}. While Eq.~\ref{eq: acceleration_eqn} depends on $d\boldsymbol{x}/d\gamma$ explicitly, there is an additional implicit dependence on this quantity through $d\boldsymbol{v}/d\gamma$ and Eq.~\ref{eq: velo}. Meanwhile, the position of the track in Eq.~\ref{eq: acceleration_eqn} is less relevant, since only its derivatives are involved when estimating accelerations. However, the position of the track is important when we need to assign a location to our estimates of the acceleration field. Because of this, we ultimately give a larger weight to the $\lambda_2$, tangent vector term in Eq.~\ref{eq: log_like_mod} due to its relative importance in estimating accelerations. However, the position of the stream track cannot be neglected, and must still remain compatible with the data. To strike a balance, we typically choose a final value of $\lambda_2$ that is a few times the order of magnitude of the mean $\lambda_1$ term in Eq.~\ref{eq: log_like_mod}. This ensures that we find a path which characterizes the mean trajectory of the stream, while also remaining compatible with the measured positions of stars along the stream. A more optimal algorithm might maximize $\lambda_2$ in Eq.~\ref{eq: log_like_mod} while simultaneously minimizing $\ell_\theta$ through the other terms in the expression. For the present analysis, however, providing a slight advantage to the $\lambda_2$ term in Eq.~\ref{eq: log_like_mod} is found to provide robust estimates of the acceleration field while remaining compatible with the data. A simple diagnostic can be performed by visually inspecting the stream track $\boldsymbol{x}_\theta(\gamma)$ compared to the position of stars along the stream. In addition, the unit vector trajectories of the stars can also be compared to the track tangent vector, $\boldsymbol{T}_\theta(\gamma)$. A successful fit will characterize the stars in both of these sub-spaces, as illustrated in Fig.~\ref{fig: Stream_Parametrized}. In this way hyper-parameter tuning is not arbitrary, since the data can be used to inform proper hyper-parameter values. Indeed, a visual inspection of the stream fits is used in this analysis when tuning hyper-parameter values. 
 
 We find that typically, $\lambda_2$ will be set to relatively large values, $\sim 10^2-10^3$, partly due to the difference in units between the $\lambda_1$ and $\lambda_2$ terms in Eq.~\ref{eq: log_like_mod} and due to the relative importance assigned to the $\lambda_2$ term. For the third penalty term in Eq.~\ref{eq: log_like_mod}, we find $\lambda_3  \sim 10$ is typically sufficient to ensure a smooth acceleration field devoid of high frequency noise artifacts from the stream track fit. We also adopt a weight decay value of $10^{-3}$ for the stream track neural network. This limits potential over-fitting. Hyper-parameter tuning can be automated using hyper parameter optimization libraries such as RayTune implemented in PyTorch \citep{liaw2018tune} to yield the best fit stream track. However, we find our analysis is robust to hyper-parameter values, provided that the individual terms in Eq.~\ref{eq: log_like_mod} receive roughly similar non-zero weight overall, with $\lambda_2$ set to somewhat larger values relative to the other two terms. Neural networks are trained for roughly 3000 epochs in mini-batches consisting of $\sim 50-100$ tracers.
 
 \subsection{Determining $\gamma$}\label{sec: gamma}
 Our modeling introduced in \S\ref{sec: technical_description} depends on a scalar $\gamma$, which acts as a phase parameter that increases monotonically along the stream. For the majority of streams discovered using e.g. {\it Gaia} so far, an on-sky position angle can be used to inform $\gamma$ since these systems do not subtend large angles, nor do they wrap back onto themselves \citep{2021ApJ...914..123I}. However, for more complicated streams like Sagittarius, an on-sky position angle is not sufficient since the stream wraps around the Milky Way several times \citep[e.g.][]{Koposov2012,Belokurov2014,2020A&A...635L...3A,2020A&A...638A.104R}. In the context of simulated data, \cite{2014MNRAS.445.3788G} devised a method to ``unwrap" a given stream using the known time evolution of each tracer. This method is not applicable to real data, for which we only have access to a kinematic snapshot of any given stream. To devise an alternative approach, we first note that there is a ``gauge freedom'' in the choice of $\gamma$, in the sense that no physically significant quantity is affected by the replacement $\gamma\to f(\gamma)$ provided the function $f(\cdot)$ is smooth and monotonic.
This can be readily seen if we consider $\gamma$ to be an on-sky position angle, e.g., $\phi_1$ in Fig.~\ref{fig: Stream_Sample}. Each tracer has its own unique angular coordinate $\phi_1$, which can be mapped to its 6d phase-space position. An equally valid mapping can be achieved in a frame which is scaled and rotated relative to Fig.~\ref{fig: Stream_Sample}, with new angular coordinates $\tilde{\phi}_1, \tilde{\phi}_2$. Consequently, $\gamma$ simply assigns an ordering to stream tracer particles, with the property that a continuous change in the parameter corresponds to a continuous change in position along the stream track.

Because of this gauge invariance, once stream tracers have been ordered, we have the freedom to assign any order-preserving scalar $\gamma$ to the stream particles. 
For complicated stream morphologies with one or several loops, we use a nearest-neighbors algorithm and an auto-encoder neural network to assign an ordering to stream stars.
We provide a technical discussion of this approach in Appendix~\ref{App: Unwrap}, and we briefly summarize the method below.

Given an unordered set of 6D phase-space measurements of stream tracers, we first construct a graph in phase-space which connects nearby particles. We impose the condition that the ray connecting neighboring segments in position space is roughly aligned with the  local trajectory of stream particles. We refer to this method as \textit{nearest neighbors with momentum}, since the algorithm moves along the stream following the local motion of tracers. This algorithm connects adjacent particles in phase-space, allowing us to assign an ordering to particles. Due to the momentum condition, the algorithm inevitably passes over some stream particles without incorporating them into the nearest neighbors graph. To fill in these gaps, we require a flexible interpolation function which maps a tracer particle in 6d phase-space to an ordered scalar parameter $\gamma$. We adopt an auto-encoder neural network to assign a suitable ordering to tracers that were left out of the graph. This neural network maps a phase-space coordinate $(\boldsymbol{x}_i, \boldsymbol{v}_i)$ to a scalar $\gamma_i$, with the arbitrary condition that the output of the auto-encoder,  $\gamma_i$, is in the interval $[-1,1]$. This method is found to successfully assign a scalar parameter $\gamma$ to a variety of stream morphologies. While this approach is not strictly necessary for simple stream morphologies, it provides an automated means to ``unwrap" a stellar stream and measure derivatives along its track. We emphasize that the unwrapping of a stream occurs as a pre-analysis step: in order to measure derivatives along the track of a stream, we must first assign an ordering to stars along the stream. For the remainder of this work, all streams are unwrapped using the automated method and the full 6D phase-space measurements for the given stream. 


We compared our nearest-neighbor-with-momentum technique to classical dimensionality-reduction methods such as kernel PCA \citep{10.5555/299094.299113} with a radial-basis-function kernel, and found that traditional techniques could not successfully unwrap stellar streams.

\section{Application to Mock Stellar Streams}\label{sec: mock_streams}

In this section we generate multiple stream systems in a ground-truth gravitational potential, and fit these streams directly using the methods introduced in \S\ref{sec: method} to sample the underlying galactic acceleration field.
\subsection{Generating Mock Stellar Streams}\label{sec: gen_mock_stream}
We focus primarily on dynamically cold streams with globular-cluster progenitors. The disruption of globular clusters has been studied extensively in direct $N$-body  simulations (see, e.g., \citealt{2003gmbp.book.....H, 2010MNRAS.401..105K}), which have shown that clusters tend to experience mass loss driven by the tidal forces of the host galaxy and two-body relaxation. Stars which escape the progenitor's tidal radius have slightly different energies from that of the progenitor, and therefore somewhat different orbits in general \citep{2011MNRAS.413.1852E,2013MNRAS.433.1813S}.

We generate mock streams in a given potential using the ``particle-spray" method from \cite{2015MNRAS.452..301F} implemented in the Gala package \citep{Price-Whelan2017}. This method is not a direct $N-$body simulation of globular cluster disruption, though it provides a prescription to reproduce streams generated in such simulations efficiently. In brief, particles are released from the progenitor near its Lagrange points for each step of an orbit integration. Released particles are then integrated forward in the background potential of the galaxy, and that of the (evaporating) progenitor system. This prescription has been shown to reproduce several detailed features of streams generated in realistic $N-$body simulations (e.g., stream fanning and morphology; \citealt{2015MNRAS.452..301F, 2015ApJ...799...28P}), and has been utilized extensively in other studies of stellar streams (e.g., \citealt{2021ApJ...910..150V,2022MNRAS.510.2437E, 2022MNRAS.tmp..239Q}).

For the galactic potential, we adopt a triaxial logarithmic potential of the form\begin{linenomath}
\begin{multline}\label{eq: log_pot}
    \Phi(x,y,z) = \\ \frac{1}{2} v^2_c \log\left[R_h^2 + \left(\frac{x}{q_1}\right)^2 + \left(\frac{y}{q_2}\right)^2 + \left(\frac{z}{q_3}\right)^2 \right].
\end{multline}\end{linenomath}
We choose this potential not because it is necessarily representative of the actual Milky Way halo. Rather, if no two of the $q_i$ are equal, the potential will not be well matched to commonly-used axisymmetric models. We adopt $q_1 = 1, \ q_2 = 1.3, \ q_3 = 0.9$, $R_h = 2$ (the lengths $R_h,x,y,z$ being measured in kpc) and $v_c = 150~\rm{km/s}$. We model the progenitor system as a Plummer-sphere potential with a mass of $2.5\times 10^4~M_{\odot}$ and a Plummer radius of $4~\rm{pc}$. We initialize the progenitor system with a randomly sampled position in a 3-dimensional box with a side length of 30~kpc. Initial velocities are sampled such that the initial speed is less than or equal to the local circular velocity.

We generate 6 streams using this method, giving rise to a diverse range of tidal debris. The progenitor's orbit is integrated forward for 6~Gyr in timesteps of $dt = 1~\rm{Myr}$, with two particles being released near the progenitor's Lagrange points at each time-step. An example of one generated stream is illustrated in the top panel of Fig.~\ref{fig: Orbit_Fitting}, in a rotated angular coordinate frame. Stream particles are colored by their energy relative to that of the progenitor.

\begin{figure}[tp!]
    \centering
    \includegraphics[scale=0.33]{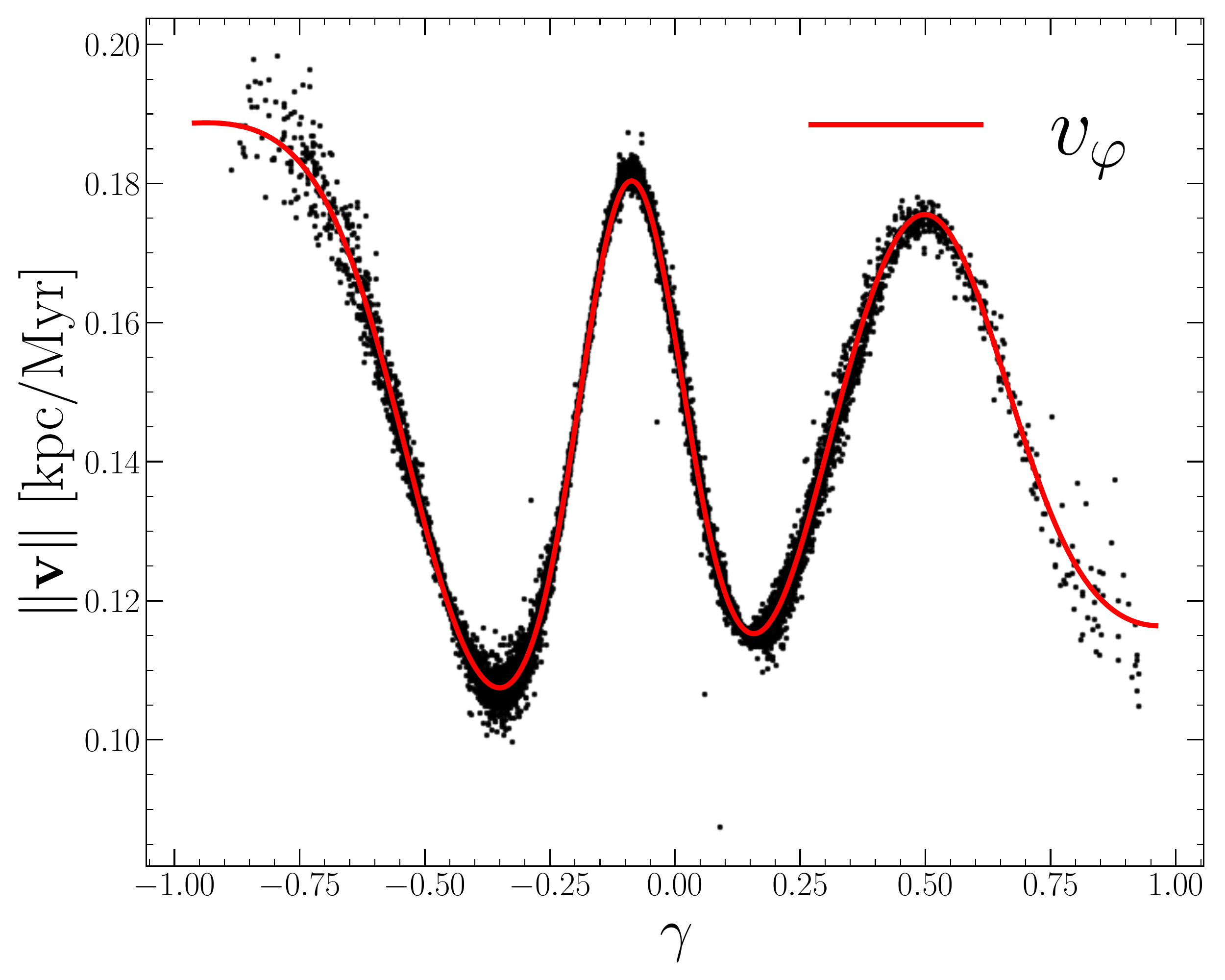}
    \caption{Black scatter points depict the speed of tracer particles for the unwrapped stellar stream illustrated in Fig.~\ref{fig: Stream_Parametrized} (top and bottom panels). The red curve is our  parametrization of the speed, fit with a neural network $v_{\varphi}(\gamma)$.  }
    \label{fig: Stream_Speed}
\end{figure}

Our method assumes that the local elongation of a given stream is tangent to the local motion of its tracers. For systems where the progenitor globular cluster is still intact, particles released from opposing lagrange points lead to sharp morphological features for which our assumptions are challenged (for example, the progenitor cluster has not fully disrupted in Fig.~\ref{fig: Stream_Sample}; $\phi_1 \approx -35.5~\rm{deg}$). We therefore remove the progenitor from our analysis on simulated data to avoid fitting this feature. For real stellar streams, only a few systems have progenitors that can still be observed (see, e.g., Pal-5, \citealt{2001ApJ...548L.165O,2017MNRAS.470...60E}; Pal-13, \citealt{2020AJ....160..244S}; Globular cluster NGC 5466, \citealt{Belokurov5466,2006ApJ...639L..17G}, and \citealt{2021ApJ...914..123I}). The progenitors of the vast majority of streams discovered so far appear to have dissolved \citep{2019ApJ...872..152I, 2019ApJ...885....3S, 2020MNRAS.494..983R, 2021ApJ...909L..26B}.  For systems with a prominent progenitor, we anticipate removing this segment of the stream to avoid bias. However, for the majority of observed streams we do not expect the progenitor to have substantial remnants. After removing progenitors, we also remove $\sim 1.5\%$ of tracers from the extended tidal tails of a given simulated stream, since it is unlikely that such extended tracers would be deemed stream members with statistical confidence. Furthermore, the quality of our derivative estimation will suffer in the limit of having only a few ($\lesssim 10$) tracers to infer dynamical changes along the stream.

These cuts are fiducial, and we find that our results are robust to small variations around these analysis choices. The typical number of stream tracers is $\sim 7000$ 
after these cuts. However, our method is applicable to streams with a significantly smaller number of measured tracers, and can interpolate over missing information. We demonstrate this advantage of our method in \S\ref{sec: orbit_fitting}, where we estimate the acceleration field along an intermediate segment of a given stream without measuring its tracers. 

\begin{figure*}[htp!]
    \centering
    \includegraphics[scale=0.27]{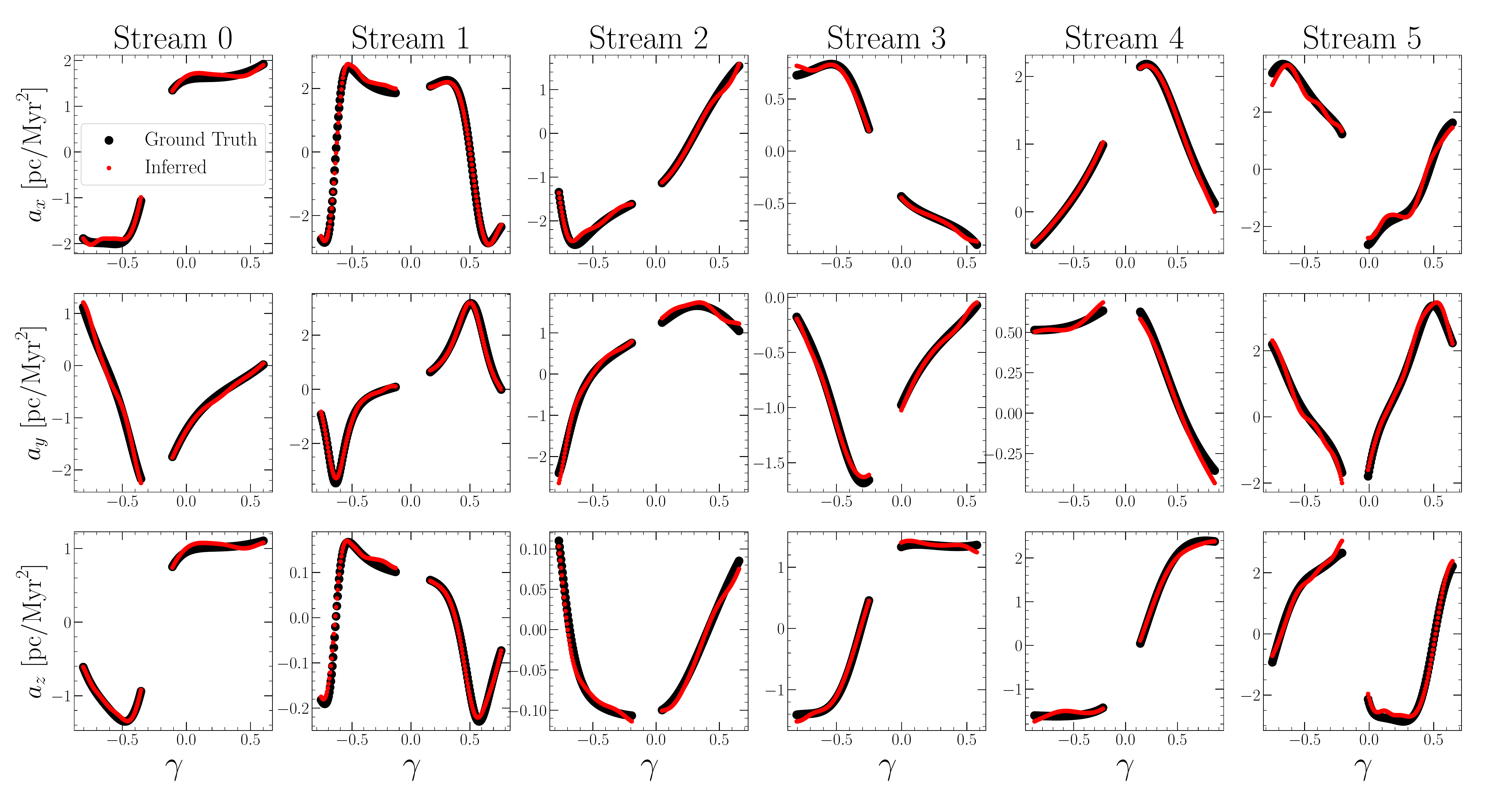}
    \caption{Reconstructed acceleration vectors along the tracks of six stellar streams. Each column is a different stream, while the 3 rows are the cartesian components of the acceleration vector. The black points depict the ground-truth acceleration at the specified position, while the red points depict our reconstruction from directly fitting the streams. Streams are parametrized in terms of the scalar parameter $\gamma$, which increases monotonically along the stream. The missing regions typically centered near $\gamma = 0$ are where the progenitor is located. We remove the progenitor from this analysis, since these regions of the stream (if observable) lead to sharp changes in morphology (see \S\ref{sec: sampling_galactic_acceleration}).  }
    \label{fig: 6_Stream_Acceleration}
\end{figure*}

\subsection{Sampling the Galactic Acceleration Field}\label{sec: sampling_galactic_acceleration}
We now apply our method introduced in \S\ref{sec: method} to six stellar streams generated in a triaxial logarithmic potential. We utilize the full 6d phase-space coordinates of stream tracers here, though we comment on the applicability of our approach to streams with missing phase-space dimensions in \S\ref{sec: future}.

In Fig.~\ref{fig: Stream_Parametrized}, we illustrate our fit to the stream track, $\boldsymbol{x}_{\theta}(\gamma)$, for a single mock stream. The fitting procedure is described in \S\ref{sec: fitting_streams}. The top row is a galactocentric view of stream tracers (black points), with the fitted track overplotted and color-coded by the $\gamma$ value for each segment of the stream. In the bottom row we plot the cartesian components of the tangent vector to the stream track, Eq.~\ref{eq: track_tangent}, in red. The black points in the bottom panels correspond to the cartesian unit-vector trajectories of individual stream tracers. Our method fits the stream track in both the space of 3d positions (top row) and 3d unit-vector trajectories (bottom row), such that the best fit track is not necessarily required to pass through the local centroid of the stream. This behavior can be seen clearly in the top row, leftmost panel around $(x,y) \approx (0,-6)$, and the top row middle panel around $(y,z)\approx (5,-4)$; in these regions, the best fit stream track is offset from the highest density of points in position space. This behavior is due to our fitting procedure, since we fit for both the position-space component of the track and its trajectory simultaneously in Eq.~\ref{eq: log_like_mod}. Consequently, the best fit path is not required to pass through the centroid of the stream; rather, the best fit track is the one that simultaneously fits the stream in the space of tracer positions (top row) and trajectories (bottom row). This fitting approach has the benefit of self-regulation, since the best fit track is the path for which our assumptions are most completely satisfied for a given stream (see \S\ref{sec: fitting_streams} for further discussion).

In order to infer a local acceleration vector along a stream, Eq.~\ref{eq: acceleration_eqn} requires a differentiable estimate of the local speed of stream tracers as a function of position along the system. As discussed in \S\ref{sec: fitting_streams}, we fit for the speed independently. In Fig.~\ref{fig: Stream_Speed}, we plot the speed of stream tracers in black as a function of the scalar parameter $\gamma$, and our fit to these points, $v_{\varphi}$, in red. Fig.~\ref{fig: Stream_Speed} corresponds to the same stream plotted in Fig.~\ref{fig: Stream_Parametrized}. Combined, Figs.~\ref{fig: Stream_Parametrized} and \ref{fig: Stream_Speed} demonstrate that a basic, fully-connected neural network is able to capture  non-trivial bulk motions of stellar tracers in a given stream, while maintaining smoothness and differentiability.  Furthermore, we note that training a neural network to generate Figs.~\ref{fig: Stream_Parametrized}, \ref{fig: Stream_Speed} is relatively fast and can be performed locally on a personal computer. 

 Once a differentiable parametrization for each stream is estimated, calculating the acceleration vector, $\boldsymbol{a}(\gamma)$, along the stream path is achieved by simply differentiating the neural networks $\boldsymbol{x}_{\theta}(\gamma)$ and $v_{\varphi}(\gamma)$ with respect to $\gamma$ through Eq.~\ref{eq: acceleration_eqn}. We emphasize that this process is performed independently for each stream. The global acceleration field can be stitched together by fitting a gravitational potential that reproduces the local acceleration field in the vicinity of each stream. In \S\ref{sec: potential}, we combine localized constraints to infer the global galactic potential. A static potential is not strictly required for the method presented in this work, provided that the analyzed stream is populated by stars with a locally similar distribution of orbits distributed over slightly different phase angles along the track of the stream. This is not the case for streams with a significant velocity component perpendicular to the track, which can occur in time-dependent potentials \citep{2019MNRAS.487.2685E}. We provide a more detailed discussion of time-dependence in \S\ref{sec: limitations}.
 
Constraints on the galactic acceleration field along six mock stellar streams are illustrated in Fig.~\ref{fig: 6_Stream_Acceleration}. Each column corresponds to a different mock-stream generated in the triaxial logarithmic potential, Eq.~\ref{eq: log_pot}. From top to bottom row, we plot the three cartesian acceleration components $a_x, \ a_y, $ and $a_z$. The black points depict the ground-truth acceleration, while the red points depict the inferred accelerations along the stream using our method. The missing gaps correspond to the removed progenitor (discussed in \S~\ref{sec: gen_mock_stream}). Across the six stellar streams in Fig.~\ref{fig: 6_Stream_Acceleration}, we find a generally
strong agreement between the reconstructed acceleration components and the ground-truth acceleration field. The typical fractional error is at the few to sub-percent level, demonstrating that our method generally provides an unbiased estimate of the local acceleration field along a given stream. In \S\ref{sec: potential}, we demonstrate that our constraints can reproduce the true parameters of the potential without bias (Fig.~\ref{fig: Analytic_potential}). 

Typically, fractional errors are largest near the ends of the streams; this is because near the boundaries of our analysis, it becomes difficult to estimate an accurate derivative for the stream track and speed in these regions. Furthermore, our neural network fits depend on many parameters $\theta$ and $\varphi$ that correspond to weights and biases for the stream track and track-speed, respectively. Consequently, the fits in Fig.~\ref{fig: 6_Stream_Acceleration} correspond to the maximum a posteriori estimates of the model parameters ${\theta}, {\varphi}$. In a future work, we intend to consider a fully Bayesian treatment of our analysis using variational inference (see  \citealt{2016arXiv160100670B} for a review), which enables us to sample from an approximate form of the posterior $P(\theta, \varphi | \mathcal{D})$. The result is a probabilistic reconstructed acceleration field, from which confidence limits can be set on galactic accelerations along stellar streams. We have carried out a preliminary test of this approach using Monte-Carlo dropout to estimate the posterior over the model parameters \citep{2015arXiv150602142G}, and find that typical deviations from the ground-truth acceleration field in Fig.~\ref{fig: 6_Stream_Acceleration} fall within the estimated model uncertainty. We postpone a detailed exploration of model uncertainty using variational methods to a future work.

We emphasize that the reconstructed acceleration field in Fig.~\ref{fig: 6_Stream_Acceleration} is the direct result of our analysis---our method does not require a model for the galactic potential to be specified as an intermediate step. This is distinct from previous analyses, which do not have this flexibility. Furthermore, we highlight that the constraints in Fig.~\ref{fig: 6_Stream_Acceleration} are not based on generative modeling; we do not rely on simplified models for the stream distribution function to constrain the acceleration field. Rather, we model the observed stream directly in a data-driven manner. We expand upon these points in \S\ref{sec: discussion}.

\section{Comparison with Orbit Fitting}\label{sec: orbit_fitting}

\begin{figure*}[htp!]
    \centering
    \includegraphics[scale=0.75]{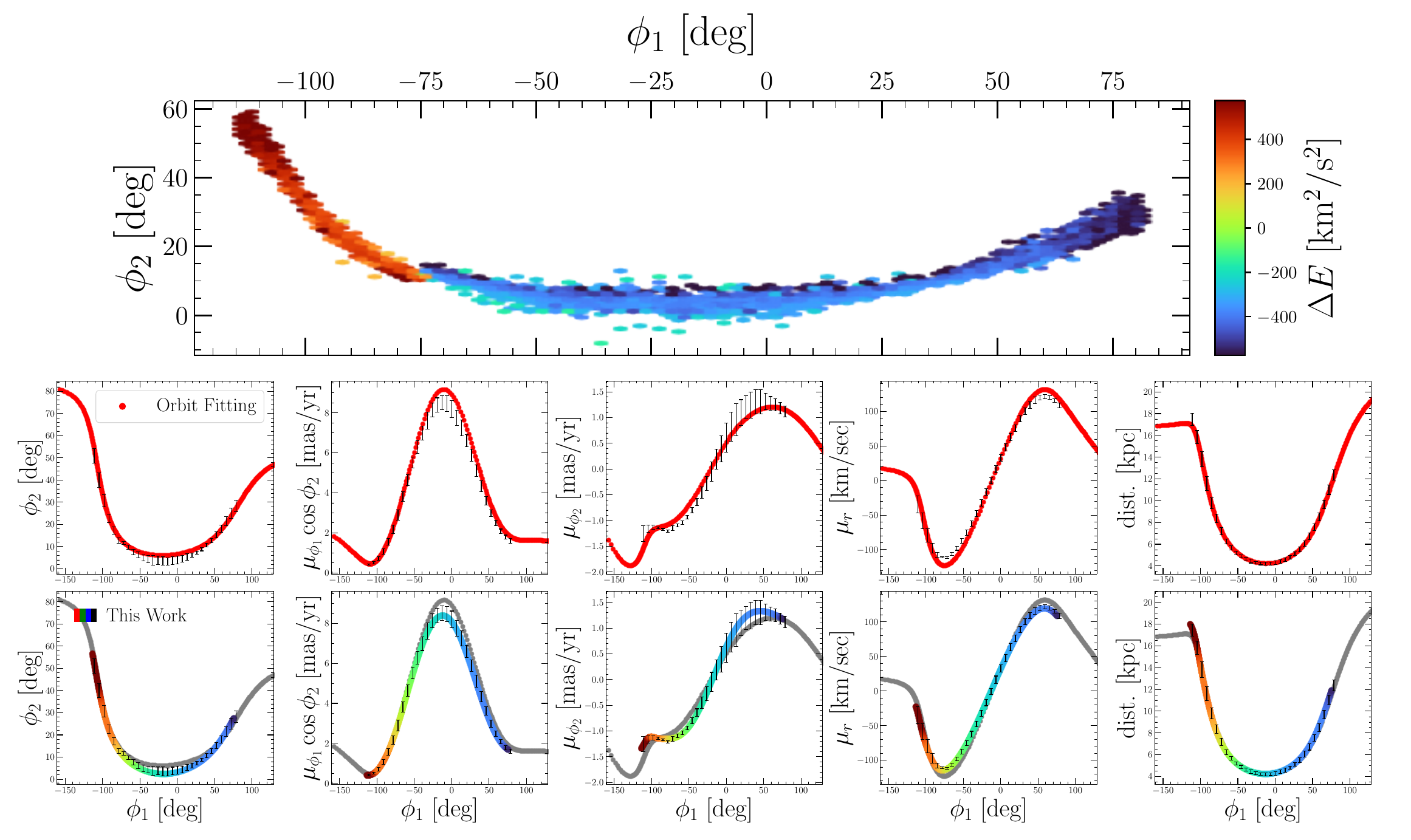}
    \caption{{\it Top Row:} Mock stellar stream generated in a triaxial logarithmic potential using the ``particle-spray" method introduced in \cite{2015MNRAS.452..301F}. $\phi_1$ and $\phi_2$ are angular coordinates which are rotated with respect to the standard equatorial frame. Stream particles are binned by their angular positions, and color-coded by the mean energy relative to the progenitor (around $\phi_1 \approx -75~\rm{deg}$). For a small band of width $\delta \phi_1$, tracers have a locally similar mixture of energies at slightly different phase-angles. {\it Middle Row:} We fit an orbit to the stream illustrated in the top row using a standard $\chi^2$ fitting routine. The binned stream measurements are illustrated by the black points with $1\sigma$ error bars, while the red curve is the best fit orbit in the logarithmic potential from Eq.~\ref{eq: log_pot}. {\it Bottom Row:} The black points with error bars and the gray curve are the same as the points and red curve in the middle panel (orbit fitting).  We overplot our stream parametrization, color-coded by the energy of the stream track in the ground-truth potential, relative to the progenitor. The color-bar for the bottom row is the same as in the top row. Our method does not require that the stream traces a path of constant energy in any potential, and captures the slowly evolving energy gradient seen among the stream in the top row.}
    \label{fig: Orbit_Fitting}
\end{figure*}
In this section we compare our method to constraints derived from orbit fitting of stellar streams. Orbit fitting has been used extensively to constrain the galactic potential or other dynamical properties from measurements of stellar streams (see, e.g., \citealt{2006MNRAS.366.1012F, 2010ApJ...712..260K, 2011MNRAS.417..198V, 2017ApJ...847..119G, 2017ApJ...836..234S, 2019MNRAS.486.2995M, 2019ApJ...881L..37B}). Typically, orbit fitting tidal streams works by first adopting a model for the galactic potential with model parameters $\Gamma$, and specifying an initial phase-space coordinate $(\boldsymbol{x}_0, \boldsymbol{v}_0)$. Trial orbits are then integrated forward from this initialization in the potential $\Phi(\boldsymbol{x} | \Gamma)$, and the parameters $\Gamma$ are adjusted between subsequent integrations until the trial orbit passes through the observed stream in phase-space. 

While appealing in its simplicity, orbit fitting assumes that streams delineate an isoenergy curve in phase-space.  In the standard context of stream formation, this cannot be the case since stars become unbound from the stream progenitor with a spread in energies \citep{1999MNRAS.307..495H,2014ApJ...795...95B,2016ASSL..420..141J}. Indeed, Fig.~\ref{fig: Stream_Sample} illustrates that stream tracers exhibit a global spread in energy and angular momentum. This can lead to a ``misalignment" between the stream track and progenitor orbit, which has been quantified in \cite{2013MNRAS.433.1813S}. 

 \begin{figure*}[tp!]
    \centering
    \includegraphics[scale=0.345]{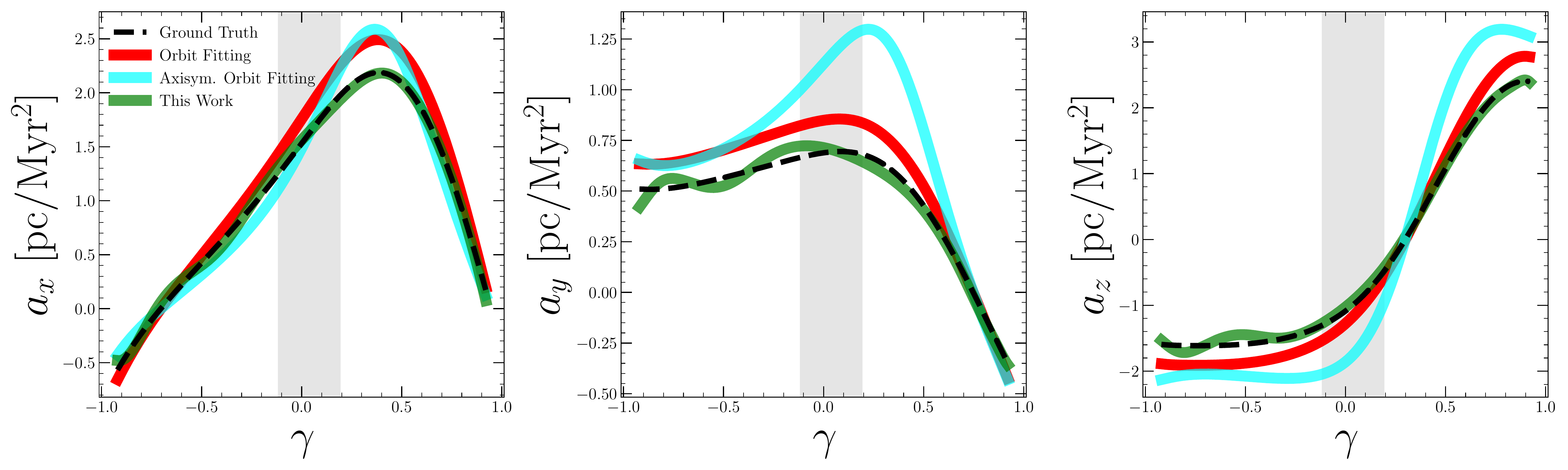}
    \caption{The inferred acceleration field in the neighborhood of the stream in Fig.~\ref{fig: Orbit_Fitting} for two methods: orbit fitting (red and cyan), our approach (green), while the ground-truth acceleration $\boldsymbol{a}(\gamma)$ is plotted in black. The red curve, derived from orbit fitting, assumes the correct model for the galactic potential. The cyan curve assumes an incorrect axisymmetric functional form for the potential. We find orbit fitting can yield a biased estimate of the acceleration field along the stream, since it incorrectly assumes that the stream is characterized by an isoenergy curve in phase-space. Adopting an imperfect model for the galactic potential is also found to yield biased estimates of the acceleration field. Our method is less restrictive, and captures the slowly evolving energy gradient among stream particles as illustrated in Fig.~\ref{fig: Orbit_Fitting} (bottom row). We also do not require an intermediate model for the galactic potential.}\label{fig: Orbit_Fitting_vs_model_free_acc}
\end{figure*}

As discussed in \S\ref{sec: method}, our approach does not require a given stream to delineate an isoenergy curve in phase-space under the correct potential, or any potential whatsoever. The phase-space coordinates of the progenitor are also not required. Additionally, our method is model independent, in that it does not require $\Phi(\boldsymbol{x}|\Gamma)$ to be specified. This differs from orbit-fitting algorithms, which require a potential model to be specified beforehand. Our approach also differs from generative methods, which forward model streams or a select number of tracers in a specified potential until a strong agreement is achieved between observational data and simulation \citep[see e.g.][]{2013MNRAS.434.2779F,2014MNRAS.445.3788G,Bowden2015,2014ApJ...795...94B, 2014ApJ...794....4P, 2018ApJ...858...73D,2019MNRAS.487.2685E, 2021ApJ...923..149S}. In addition to requiring a parametric model for the potential, generative methods often require a reliable, efficient prescription for stream formation. Our method does not require these assumptions or modeling capabilities. Instead, the approach introduced in this work assumes that contiguous segments of the stream are composed of stars with a similar mixture of energies. Hence the change in energy along the stream is small, though not necessarily zero. 

We next illustrate the bias induced by assuming that stellar streams trace orbits in the galactic potential, and how this bias is alleviated with our approach. We also illustrate bias in estimated accelerations as a result of adopting a slightly incorrect functional form for the galactic potential. To accomplish these experiments, we use the stream illustrated in the top row of Fig~\ref{fig: Orbit_Fitting}. This stream is typical in our simulations, and its progenitor is clearly still prominent. In real data, such a stream would be a strong candidate for orbit fitting, since the present day phase-space coordinates of the progenitor can be integrated forward and backward in a specified potential, adjusting model parameters until the orbit passes well through the stream. We carry out this exercise on the synthetic stream in Fig.~\ref{fig: Orbit_Fitting} (top row), using the current phase-space coordinates of the progenitor and 25 other stars at the trailing tail of the stream as trial initial conditions.

Given that we generated the streams used in this analysis, we are at a distinct advantage of choosing an accurate model for the galactic potential. Of course, on real data it is unclear what parametrization should be adopted. We therefore integrate trial orbits in two potentials. The first is the correct model for the potential, Eq.~\ref{eq: log_pot}, with model parameters $\Gamma = (v_c, r_h, q_1, q_2, q_3)$. The second is an axisymmetric version of the same potential, with fixed parameters $q_1=q_2=1$. We allow $q_3$ to remain a free parameter, characterizing the z-axis flattening. This is a common choice for the potential in previous works, which have applied orbit fitting to real stellar streams (e.g., \citealt{2010ApJ...712..260K}). Trial orbits are integrated in the two potential models separately, and compared to binned stream measurements in a rotated ICRS frame. Model parameters are adjusted until the $\chi^2$ between the observed stream and integrated orbit is minimized for each initial condition using a \texttt{Scipy} \citep{2020SciPy-NMeth} optimizer. This approach is similar to \cite{2010ApJ...712..260K}, where orbit fitting is used to constrain a potential model from the GD-1 stream. The best fit orbit for the stream in Fig.~\ref{fig: Orbit_Fitting} (top panel) using the correct model for the galactic potential is illustrated in the middle row of Fig.~\ref{fig: Orbit_Fitting} in red, where the orbit is plotted in a rotated ICRS frame. The quantities $\mu_{\phi_1}, \mu_{\phi_2}, \mu_r, \rm{and \ dist.}$ indicate the two proper motion components, radial velocity, and distance along the orbit, respectively. Binned stream measurements are shown by the black points with error-bars, where the error-bars are derived from the intrinsic scatter of stream observables in each $\phi_1$ bin.

 We find that the 5-parameter potential, Eq.~\ref{eq: log_like_mod}, is flexible enough to generate an orbit which passes through the binned stream measurements. However, because the stream does not actually delineate an isoenergy curve, the resulting best fit orbit and corresponding potential yield estimates of the acceleration field which are not necessarily accurate. We illustrate this bias in Fig.~\ref{fig: Orbit_Fitting_vs_model_free_acc}, where we compare our inference of the acceleration field along the stream (green) to orbit fitting (red) against the ground-truth acceleration (dashed black curve) for the cartesian components $\boldsymbol{a} = (a_x, a_y, a_z)$. The cyan curve is the result of integrating trial orbits in an incorrect axisymmetric model for the galactic potential as previously described. We include this curve to illustrate bias induced by adopting an incorrect functional form for the potential. Both the red and cyan orbit-fitting curves are taken from the negative spatial gradient of the best-fit potential corresponding to the best-fit trial orbits (e.g., Fig.~\ref{fig: Orbit_Fitting} middle row). Estimates of enclosed mass at the progenitor's location for the best fit potential derived from orbit fitting is biased from the truth by roughly 10\% when using the correct potential model, and up to 25\% when imposing axisymmetry. For the method presented in this work, we find fractional errors on enclosed mass at the sub-percent ($\sim 0.9\%$) level for the dashed black curve in Fig~\ref{fig: Orbit_Fitting_vs_model_free_acc}.   
 
 Contrasted with orbit fitting, our method allows for the possibility that globally, streams depart from a single orbit. Indeed, for most regions of the stream in Fig.~\ref{fig: Orbit_Fitting_vs_model_free_acc}, the true acceleration components and the components inferred from our method are similar. Furthermore, our method does not require priors on the functional form of the potential: this enables increased flexibility and significantly reduces systematic errors from adopting an incorrect model for the galactic potential. Fluctuations in the inferred accelerations are attributed to model uncertainty in our neural network based-fits, which we intend to quantify in a future work using variational inference. We also note that for some streams analyzed in this work, orbit fitting can provide unbiased estimates of the acceleration field along the given stream under the correct model for the potential. However, all streams yield biased constraints on local accelerations under the incorrect functional form for the galactic potential. Substantially increasing the flexibility of the potential by adding more model parameters is one possible solution, though this is also found to yield noisy estimates of the local acceleration field when using orbit-fitting techniques. 
 
 In Fig.~\ref{fig: Orbit_Fitting}, bottom row, we show our parametrization of the stream track in the same rotated ICRS frame, with the best fit orbit from the top row plotted in grey, and the neural-network-derived stream track illustrated by the multi-colored curve. The stream track is color-coded by its energy in the correct underlying potential (relative to the progenitor). We find an energy gradient, which closely follows the mean relative energy of tracers in the top row of Fig.~\ref{fig: Orbit_Fitting} (note that the color-scale in the top row of Fig.~\ref{fig: Orbit_Fitting} and bottom row are the same). Taken jointly, Fig.~\ref{fig: Orbit_Fitting} and Fig.~\ref{fig: Orbit_Fitting_vs_model_free_acc} illustrate that our approach is significantly less restrictive than orbit-fitting, since we recover the local energy gradient of stream tracers (Fig.~\ref{fig: Orbit_Fitting}, bottom row) while also recovering the correct acceleration field along the stream (Fig.~\ref{fig: Orbit_Fitting_vs_model_free_acc}, green curve). Both of these results are achieved without having to model the gravitational potential directly. This is advantageous, since inaccurate models for the potential can lead to bias, even for less restrictive potential reconstruction methods based on generative modeling. 
 
 We also use Fig.~\ref{fig: Orbit_Fitting_vs_model_free_acc} to illustrate that our analysis can be applied to streams with missing observations. For all streams analyzed using our method in this work, we remove the progenitor system (see \S\ref{sec: fitting_streams} for discussion). When estimating the acceleration field in Fig.~\ref{fig: Orbit_Fitting_vs_model_free_acc}, we do not rely on any tracers that fall within the grey band of each panel since this region is roughly the location of the progenitor cluster. This corresponds to the tracers centered around $\phi_1\approx -75~\rm{deg}$ in the top row of Fig.~\ref{fig: Orbit_Fitting}. We find that our method is capable of flexibly interpolating over this missing information. This is useful for real stellar streams, since the observed tracers are generated from both the stream distribution function and the observational selection function; our method does not require these distributions to be disentangled. 
 
\section{Potential Reconstruction}\label{sec: potential}
From a Fisher-information analysis of stellar streams, \cite{2018ApJ...867..101B} found that streams provide a highly localized measurement of the galactic halo. Because of this, they suggest fitting streams independently, and combining constraints on the galactic potential at the global level as a separate step. Our method provides the framework to accomplish this, since we estimate the local acceleration field along each stream rather than global properties of the halo. In this section, we explore how our local estimates of the galactic acceleration field can be combined to constrain a global gravitational potential. 

We consider a static potential, though will explore time-dependent potentials in a future work. In principle, a static potential is not strictly required in our analysis. The central requirement of our work is that the stream forms a coherent phase-space object, with the property that adjacent segments of the stream are populated by stars with a locally similar distribution of orbits. For streams with a strong misalignment between the stream-track and stellar trajectories, this assumption is broken. A track-velocity misalignment has been shown to occur in time-dependent potentials with (e.g.) a Large Magellanic Cloud component \citep{2019MNRAS.487.2685E}. Still, depending on the magnitude and evolution of the misalignment across the stream our analysis could still be applicable. We provide further discussion of a time-dependent potential in \S\ref{sec: future} and discuss extensions to the present work. 

The gravitational potential is related to the acceleration field through Eq.~\ref{eq: a_nabla_phi}. This relation can be used to find the potential, $\Phi(\boldsymbol{x})$, which gives rise to the local acceleration field in the neighborhood of independent streams. Fitting a model for the potential in this framework will typically be computationally inexpensive, provided that we have already estimated the acceleration field along independent streams using the method introduced in this work. This is unique and distinct from previous analyses, which constrain the galactic potential or other dynamical properties of the Milky Way using generative methods and several forward simulations (see, e.g., \citealt{2013MNRAS.434.2779F,2014ApJ...795...94B, 2014ApJ...794....4P,2014MNRAS.445.3788G,Bowden2015,2015ApJ...799...28P, 2018ApJ...858...73D, 2019MNRAS.487.2685E,2021ApJ...923..149S}). These works typically fix an analytic model for the potential with parameters free to vary. However, choosing a model in the first place imposes a functional prior on the potential. That is, a given potential model is never guaranteed to describe the actual observed galaxy. Our method alleviates the need to specify a potential model beforehand, letting the data drive our inference. However, priors on the functional form of the potential can still be adopted (e.g. by assuming that $\boldsymbol{a}$ can only depend on galactocentric distance $r$). 

Using the methods outlined in this work, a range of analytic models for the potential can be constrained quickly using standard optimization algorithms, enabling model comparison techniques which would be too expensive under the generative framework. Alternatively, a highly-flexible representation for the potential can be constrained using (e.g.) basis function expansions or a flexible neural network to represent $\Phi$. This has the advantage of preserving the detailed features of the acceleration field around each measured stream, while interpolating between independent systems. Provided that the data have enough constraining power, flexible models minimize the need to construct an accurate analytic form for the potential beforehand. 

We therefore consider two pathways for potential reconstruction. The first fits a standard analytic model for the potential to our estimate of the acceleration field along the six stellar streams analyzed in \S\ref{sec: sampling_galactic_acceleration}. This enables one to derive constraints on the parameters of the model for the Galactic potential. The second uses a neural network to describe the potential that gives rise to the estimated acceleration field in the vicinity of each stream, enabling a flexible stream-based analysis of the galactic halo that is fully data-driven. 

The output of our main analysis in \S\ref{sec: sampling_galactic_acceleration} are acceleration components sampled across several streams. In the subsequent sections, we denote the set of these estimates with $\{\boldsymbol{a}(\boldsymbol{x}_i)\}_i$, where $\boldsymbol{x}_i$ is the $i^{\rm th}$ position at which the acceleration field is evaluated.

\subsection{Constraining an Analytic Potential}\label{sec: analytic_potential}
In this section we fit an analytic model for the potential to our estimate of the  acceleration field along the six independent streams in \S\ref{sec: sampling_galactic_acceleration}. A wide range of parametric models have been used to describe the Galactic potential. Choosing the wrong model can lead to bias (\citealt{2014ApJ...795...94B}, and our Fig.~\ref{fig: Orbit_Fitting_vs_model_free_acc}), making model selection an important step in constraining the potential. For real data, it is not obvious what model should be adopted beforehand. On simulated data we are provided with a distinct advantage: we know the correct analytic form of the potential. Therefore, in this section we adopt the correct analytic form of the ground-truth potential, Eq.~\ref{eq: log_pot}, and allow its parameters to vary. However, we emphasize that our analysis enables the use of model comparison techniques without the added computational cost of simulation based inference or generative modeling. A detailed statistical model comparison of the Milky Way potential using stellar streams and e.g. the Bayesian evidence framework has not yet been carried out. This is likely due to the large computational cost of generating a new set of simulations for each model and its corresponding parameter grid. Because our analysis constrains accelerations rather than the potential, it enables us to compare analytic potential fits rapidly. We intend to explore this aspect of our method in a future work with data from real stellar streams.

 \begin{figure}[tp!]
    \centering
    \includegraphics[scale=0.32]{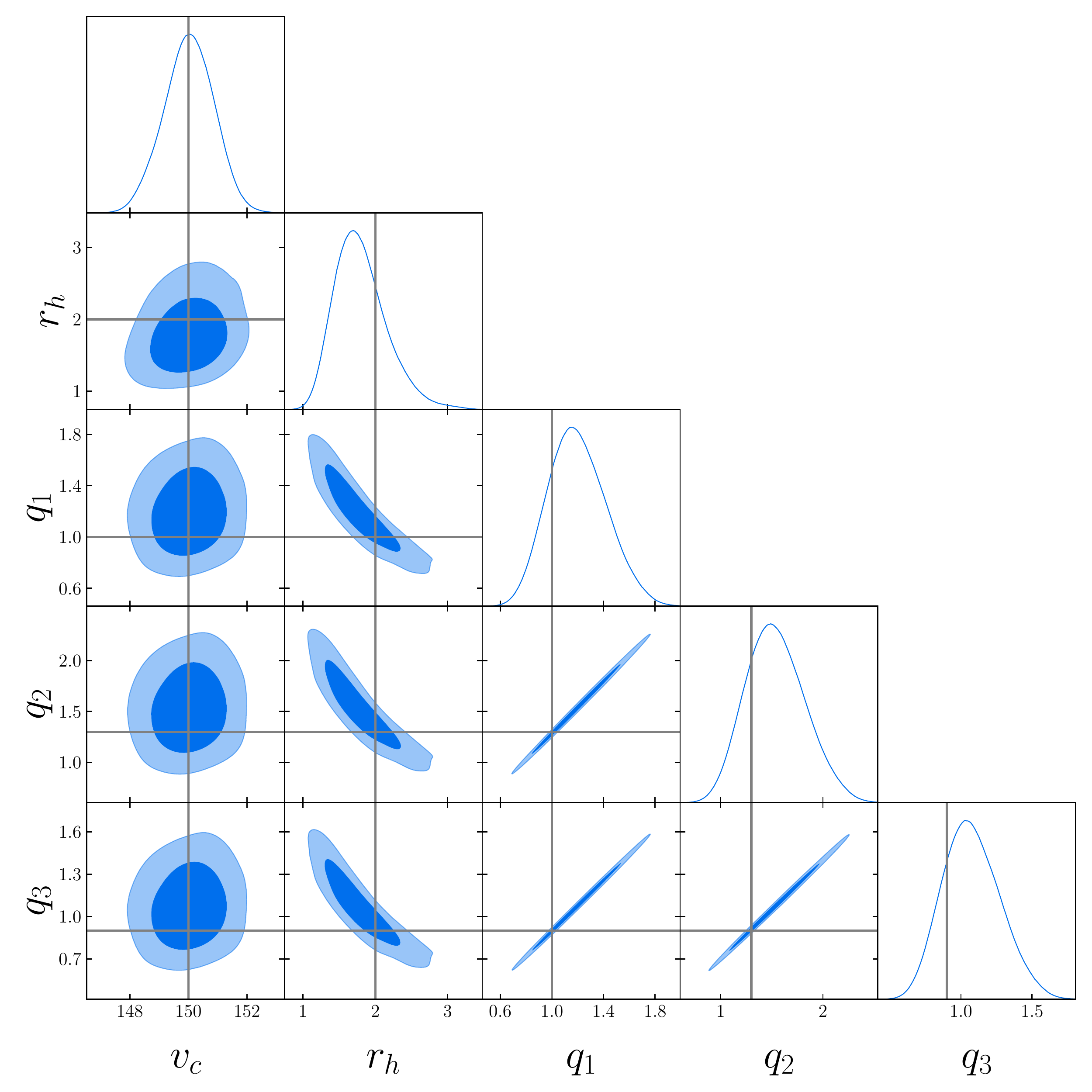}
    \caption{Constraints on model parameters for the logarithmic potential in Eq.~\ref{eq: log_pot}. Parameter constraints are derived from the estimated acceleration field along the six mock-streams analyzed in \S\ref{sec: sampling_galactic_acceleration}. Contours correspond to regions of 68 and 95\% confidence, and gray lines depict the true underlying parameter values. }\label{fig: Analytic_potential}
\end{figure}

We constrain the logarithmic potential in Eq.~\ref{eq: log_pot} with the five parameters $\Gamma = \left(v_c, r_h, q_1, q_2, q_3\right)$ using a \texttt{Scipy} \citep{2020SciPy-NMeth} non-linear least-squares optimizer. In particular, we minimize the $\chi^2$
\begin{equation}\label{eq: chi2}
    \chi^2 = \sum_i \frac{\Vert \boldsymbol{a}\left(\boldsymbol{x}_i\right) + \nabla \Phi\left(\boldsymbol{x}_i|\Gamma\right)\Vert^2}{\sigma_i^2},
\end{equation}
over the parameters $\Gamma$, where $\sigma^2_i$ are the estimated uncertainties on the inferred acceleration field at each evaluation point $\boldsymbol{x}_i$. In \S\ref{sec: sampling_galactic_acceleration}, we discuss how variational inference can be used to obtain an estimate of this uncertainty, though we set all $\sigma_i \equiv 1$ for now. We estimate errors on the model parameters $\Gamma$ by boot-strap resampling the inferred acceleration field along each stream. In particular, we minimize Eq.~\ref{eq: chi2} by sampling the acceleration field at 100 random positions across all streams. We repeat this many times, minimizing Eq.~\ref{eq: chi2} for each realization to estimate errors on the best fit model parameters. We impose the constraints $v_c \in [50,400], \ r_h \in [.2,10], \ q_i \in [0.1,5]$ for $i = 1, 2, 3$. These constraints are not strictly necessarily, though aid in optimization since the logarithmic potential in Eq.~\ref{eq: log_pot} is degenerate in $r_h$ and the flattening parameters $q_1,q_2,q_3$. That is, if one replaces $r_h$ with $\alpha^{-1}r_h$ and $(q_1,q_2,q_3) \to (\alpha q_1,\alpha q_2, \alpha q_3)$ in the potential (Eq.~\ref{eq: log_pot}) for any constant scale factor $\alpha >0$, then $\Phi$ changes by an additive constant. However, the underlying accelerations remain unchanged. 

\begin{figure*}[tp!]
    \centering
    \includegraphics[scale=0.65]{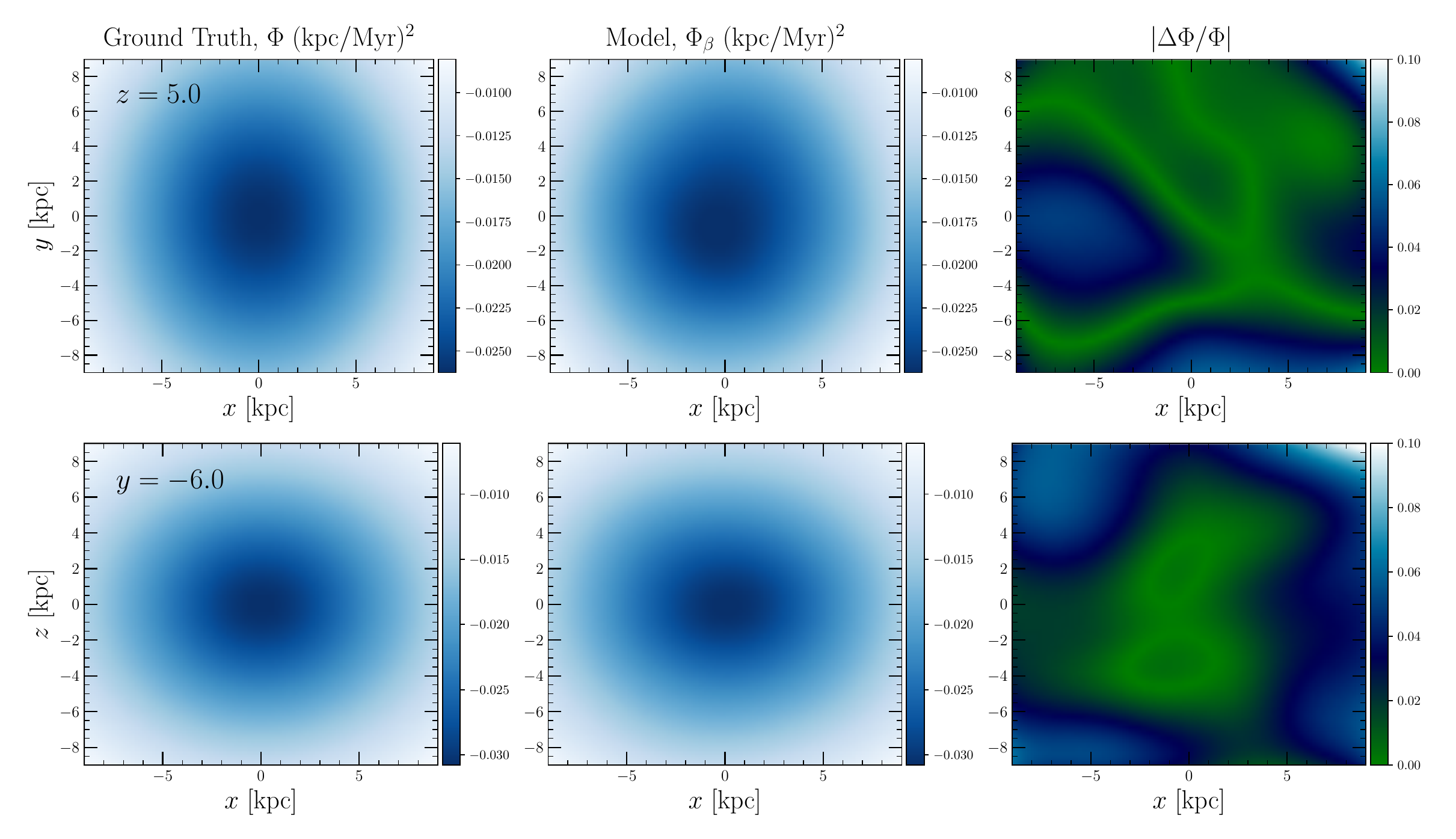}
    \caption{Flexible potential reconstruction using six stellar streams and their inferred acceleration field, illustrated in Fig.~\ref{fig: 6_Stream_Acceleration}. {\it Top row:} The left panel is the ground truth potential in a galactocentric frame with $z$ fixed to 5~\rm{kpc}. The middle panel is the flexible reconstruction, and the right panel illustrates fractional errors.
    {\it Bottom row:} Same as the top row, but for $y = -6~\rm{kpc}$.}\label{fig: Reconstructed_Pot}
\end{figure*}

The resulting distribution of best-fit parameters is illustrated in Fig.~\ref{fig: Analytic_potential}, where grey lines indicate the true parameter values and contours correspond to regions of $68$ and $95\%$ confidence. From Fig.~\ref{fig: Analytic_potential}, we find that our method for estimating the local acceleration field along independent streams provides combined parameter constraints which fall within the $68\%$ region of the true potential parameters. Adopting different analytic models for the potential $\Phi$ in Eq.~\ref{eq: chi2} is straightforward to implement, since our estimate of the acceleration vector, $\boldsymbol{a}$, along each stream is independent of any particular parametrization of the potential. 

\subsection{Constraining a Flexible Potential}\label{sec: flexible_potential}
The true Galactic potential and its dark halo are not necessarily well captured by simplified analytic (e.g. spherical) models \citep{2006MNRAS.367.1781A,2009ApJ...703L..67L,2011MNRAS.416.1377V,2015ApJ...805...29R,2020ApJ...893...15A,2022arXiv220103479B,2021ApJ...919..109G}.  Therefore, model selection becomes an important step in constraining the galactic and halo potential from stellar streams. Provided that the data have strong constraining power, flexible models for the gravitational potential have the ability to capture a diverse range of features across the galaxy while reducing the bias induced by adopting the wrong parametric model. Basis-function expansions provide one pathway to parametrizing a flexible potential, and can be fit with standard least-squares optimization routines or MCMC. 

We consider an alternative approach, which uses a flexible neural network to describe the Galactic potential. This has the advantage of increased flexibility over basis-function expansions. In particular, a basis and truncation condition do not need to be specified. However, we emphasize that basis expansions can equally be applied to constrain a potential from our estimates of the acceleration field. Following the approach of \cite{2020arXiv201104673G, 2022arXiv220502244G}, we represent the potential $\Phi$ with a flexible fully-connected neural network $\Phi_{\beta}(\boldsymbol{x})$, where $\beta$ are the parameters of the neural network. Using estimates of the acceleration field localized to several streams, $\Phi_{\beta}$ can be trained through its spatial gradient to reproduce the local acceleration field of stellar streams while interpolating over the unsampled regions. Poisson's equation connects the gravitational potential to the matter density,
\begin{equation}\label{eq: poisson}
   \nabla_{\boldsymbol{x}}^2 \Phi = 4\pi G\rho, 
\end{equation}
where $G$ is the gravitational constant and $\rho(\boldsymbol{x})$ is the matter density of the Galaxy. In order to ensure that the inferred potential is physical, we would like to encourage the condition 
\begin{equation}\label{eq: poisson_2}
    \nabla^2\Phi_{\beta}(\boldsymbol{x}) \geq 0
\end{equation}
to promote a non-negative mass density. To accomplish this, we use a loss function that penalizes negative mass densities during training. The loss function is\begin{linenomath}
\begin{multline}\label{eq: potential_loss}
        \ell_{\beta}(\beta)
         \equiv \frac{1}{N}\sum_{i=1}^N \Bigg[ \left(\nabla_{\boldsymbol{x}}\Phi_{\beta} \Big\vert_{\boldsymbol{x}_i} + \boldsymbol{a}(\boldsymbol{x}_i) \right)^2 \\ + \lambda\,\texttt{max}\left(0, -\nabla^2_{\boldsymbol{x}} \Phi_{\beta} \Big\vert_{\boldsymbol{x}^*_i}\right)\Bigg],
\end{multline}\end{linenomath}
where $\boldsymbol{x}^*_i$ is sampled within the region of interest (e.g. a 3-dimensional box within which all streams are contained).  The $\lambda$ term in Eq.~\ref{eq: potential_loss} works to penalize randomly sampled regions with a negative mass density. In practice, we also sample positions along each stream. To encourage smoothness, weight decay is also used to penalize large neural-network parameters.  We use a weight-decay value of $\sim 10^{-5}$. We apply this method to the six stellar streams whose local acceleration fields are inferred in Fig.~\ref{fig: 6_Stream_Acceleration}, with $\lambda = [0.05, 0.5, 1]$ after $\sim$ 200, 500, and 800 training epochs, respectively. We illustrate the reconstructed potential in a Galactocentric view in Fig.~\ref{fig: Reconstructed_Pot} for two different slices: the top row shows a constant $z$ slice through our mock galaxy, while the bottom row shows a slice with constant $y$. The first column depicts the ground truth potential (Eq.~\ref{eq: log_pot}), the middle column is the inferred potential, $\Phi_\beta$, and the right column depicts fractional errors. We note that the color maps in the first two columns have the same range, allowing for a visual comparison between the true potential and our inference. Across a range of scales, we recover an accurate depiction of the potential with fractional errors typically at the few-percent to sub-percent level.

 The only imposed constraint in our inference of the potential in this section is the non-negativity of mass. The neural network $\Phi_{\beta}$ knows nothing of the triaxility of the potential beforehand: it discovers this from the inferred acceleration field directly. Using the acceleration field sampled from only six stellar streams (Fig.~\ref{fig: 6_Stream_Acceleration}), our flexible neural-network-based reconstruction reproduces the features of the global triaxial potential across a range of scales, even in intermediate regions where mock-streams do not reside. We find the most significant bias at $\boldsymbol{z} = 0$ near the galactic center, since streams were not generated in this vicinity; they would quickly disrupt anyway due to extreme tidal forces. 
 
The residuals in Fig.~\ref{fig: Reconstructed_Pot} are quite structured, especially in the $xy$-plane. Fractional errors tend to increase away from the six streams, leading to the stream-like structure in the residuals of Fig.~\ref{fig: Reconstructed_Pot}. This is in agreement with \cite{2018ApJ...867..101B}, who found that fitting streams under flexible potential models constrains the galactic acceleration field localized to the stream(s) of interest. The pericenters of the six streams analyzed fall between $[4.5,8]~\rm{kpc}$, with apocenters between $[7.5,16]~\rm{kpc}$. Beyond $\sim 16~\rm{kpc}$ fractional errors increase.    
 
 It appears somewhat surprising that the neural network-based potential can accurately interpolate over unsampled regions of the galactic acceleration field. We aim to explain this in the following way. When fitting an over-parametrized linear regression model using gradient descent, one implicitly optimizes for the smoothest possible solution that is also compatible with the data \citep{TheoryOfDeepLearning}. In particular, the optimal parameter set is the pseudoinverse solution, which has the smallest $L_2$ norm over the model parameters. As an analogous system, we expect that the same is true for deep neural networks trained with gradient descent. Because the $L_2$ norm over neural-network parameters is directly correlated with the Lipschitz constant of the network (defined as $\rm{sup}_x[f^\prime(x)]$; \citealt{2018arXiv180404368G}), the optimal neural-network parameters also implicitly minimize the upper bound of this constant. In addition, weight decay explicitly minimizes the $L_2$ norm over the parameters. The result is that neural-network-based regression yields the smoothest possible solution for which the upper bound of the Lipschitz constant is minimized while also maintaining compatibility with the data. For this reason, along with the enforcement of a potential with a non-negative mass density, it is actually not unusual that the best-fit neural-network potential smoothly interpolates over regions without measured accelerations. This is expected for a neural network that characterizes the measured portions of the acceleration field, while also optimizing for smoothness. 
 
 The advantage of the neural-network-based potential presented in this section is that one does not need to specify an analytic model for the potential. In this way, a neural-network description enables potential reconstruction from a fully data-driven standpoint. \citet{2018ApJ...867..101B} demonstrate that flexible models for the potential characterize the local acceleration field in the vicinity of a given stream. A neural-network-based model is flexible enough to fit a potential that precisely reproduces the local acceleration field inferred from independent streams, while also characterizing the potential at a global scale. 
  This is enabled by increased flexibility, and the ability to minimize non-physical artifacts through Eq.~\ref{eq: poisson_2}. Flexible potential models come at the cost of reduced model interpretability, though we intend to explore symbolic regression (e.g., \citealt{2020arXiv200611287C}) applied to our flexible inference of the potential in a future work. 
 
\section{Discussion}\label{sec: discussion}
\subsection{Summary}
We have introduced a new method to constrain the galactic acceleration field from measurements of stellar streams. Our method is agnostic to the functional form of the underlying potential and does not require streams to delineate orbits. Rather, our central assumption is that streams are populated by an ensemble of stars characterized by a local mixture of orbits that changes gradually along the stream. Therefore, our approach does not require streams to delineate isoenergy curves in phase-space. Rather than relying on generative models and simulation based inference, we fit the dynamics of the bulk motion of the observed stream directly as a function of position along the stream track. Using a flexible neural network parametrization of the stream track, derivatives of dynamical quantities (e.g. position and velocity) can be calculated, providing a direct link to the local acceleration field in which the stream resides. Our method does not require approximate transformations to action-angle coordinates, but has the benefit of operating in phase-space. 

Our method for sampling the galactic acceleration field with stellar streams is methodologically different from previous works, which typically rely on orbit fitting, generative models, numerous orbit integrations, or action-angle coordinate transformations to constrain an analytic potential model using stellar streams. We have compared our approach to orbit fitting in \S\ref{sec: orbit_fitting}, where we demonstrated that our approach alleviates inaccuracies in the inferred accelerations induced by assuming that streams delineate isoenergy curves in phase-space. Because our approach estimates accelerations directly without the intermediate requirement of specifying a model for the potential, we also alleviate bias induced by adopting an incorrect model for the potential (Fig.~\ref{fig: Orbit_Fitting_vs_model_free_acc}). This enables a description of the acceleration field which is independent of any user-defined functional form for the potential, allowing us to accurately probe halo morphologies with complex or unexpected geometries.  Furthermore, we do not require a generative prescription for stream formation, nor do we require knowledge of e.g. the present day phase-space coordinates of the stream progenitor. Instead, we assume that small neighboring segments of the stream are populated by stars with a similar distribution of orbits. This makes our analysis applicable to e.g. {\it Gaia}-detected stellar streams for which the progenitor has completely disrupted, or streams which are not well modeled in current generative frameworks. 

\cite{2018ApJ...867..101B} suggested that stellar streams probe highly localized properties of the underlying galactic potential; our analysis exploits this as a feature of stellar streams to constrain the galactic acceleration field on small scales. As a post-processing step, we demonstrate that highly localized measurements can be combined across several streams to place constraints on a global gravitational potential. Indeed, we find that our method is able to reproduce the correct parameters of a global ground truth static potential without bias (\S\ref{sec: analytic_potential}). We have also demonstrated that a highly flexible potential model can be constrained, enabling potential reconstruction from a fully data-driven standpoint (\S\ref{sec: flexible_potential}).  For more complicated potentials beyond a triaxial model, the analysis presented in this work has the benefit of flexibility and does not require prior knowledge of the functional form for the gravitational potential. We intend to explore this benefit of our analysis in further detail, through an application to cosmological simulations and real data of Milky Way streams.

\subsection{Limitations}\label{sec: limitations}
The Milky Way is currently undergoing a major merger with the Large Magellanic Cloud (LMC; e.g. \citealt{2007ApJ...668..949B,2016MNRAS.456L..54P,Erkal2020bias,Erkal2021sloshing,2021Natur.592..534C}). For streams in the Galactic Southern hemisphere in particular, the LMC is likely to impart time-dependent perturbations on the morphology and kinematics of these systems \citep{2015MNRAS.450.1136E, 2021ApJ...923..149S}. Indeed, recent work has explored the dynamical influence of the LMC on the Tucana, Orphan and Sagittarius streams, since these systems may have suffered a direct interaction or close passage with the LMC \citep{2018MNRAS.481.3148E,2019MNRAS.485.4726K,2021MNRAS.501.2279V, 2022arXiv220501688L}. Direct interactions with the LMC can lead to a significant misalignment between the stream track and proper motions, which has been shown to occur only in the presence of time-dependent perturbations \citep{2019MNRAS.487.2685E}. This misalignment has been used to measure properties of the LMC such as its mass \citep{2019MNRAS.487.2685E,2021ApJ...923..149S}, but poses a potential challenge in applying our analysis to these systems. The method introduced in this paper requires the local motion of stream stars to roughly fall along the stream track. When this assumption is broken, our method in its current formulation is no longer directly applicable. However, because our analysis has the benefit of self-regulation (see \S\ref{sec: method}), it will be clear when a given stream is a poor candidate for our approach. In a future work, we intend to consider a more general formulation of our methodology which will be applicable to streams which have been influenced by time-dependent perturbations, and therefore have a misalignment between proper motions and the stream track. We discuss one possible formulation in \S\ref{sec: future}. Still, there are a number of streams which do not appear to have suffered from time-dependent perturbations (e.g. see the ``golden sample" used in \citealt{2020arXiv201205271M}). We therefore 
anticipate that several streams can be explored using the methods presented in this work, particularly in the galactic Northern hemisphere. 

Recent work has detected a relative velocity difference between the outer Milky Way halo ($r \in [40,120]~\rm{kpc}$) and the galactic disk, thought to be a result of the Milky Way's dynamical response to the LMC \citep{2019ApJ...884...51G,Erkal2020bias,2020MNRAS.494L..11P,Erkal2021sloshing}. This could lead to a velocity misalignment between the stream track and local motion, as a result of the non-inertial reference frame of the disk. \cite{2020arXiv201205271M} has measured the misalignment between the proper motions of stream stars and stream track for streams within $\sim 25~\rm{kpc}$ of the galactic center, and found that the misalignment was in agreement with the solar reflex velocity. This indicates that the inner halo may have a negligible relative velocity compared to the galactic disk. We therefore anticipate that our analysis will be less sensitive to systematic errors induced by the Milky Way's reflex motion for streams within a $\sim 25~\rm{kpc}$ volume.

Our method will be most applicable to static halo potentials, though this condition is not strictly required. In particular, our core assumption is that streams contain stars with a locally similar mixture of orbits at slightly different phase angles (Fig.~\ref{fig: Stream_Sample}). Provided that the potential is evolving slowly, integrals of motion (e.g. actions) will remain adiabatically invariant so that this assumption will not necessarily be broken. Even for more strongly time-dependent halo potentials in which orbital deviations are non-adiabatic, orbital actions may remain locally comparable so that we can still treat streams as locally similar mixtures of stellar orbits  \citep{2015A&A...584A.120B,2021MNRAS.501.2279V}. Because our method for estimating galactic accelerations works by measuring changes in local dynamical properties along a stream, streams with a high number density of stars and ample phase-space observations will provide more accurate measurements of the galactic acceleration field over recent timescales. That is, if we view the mean stream track as an ensemble of several distinct orbital segments, then the length of each measurable segment becomes smaller as the number of measurable stream stars increases. Provided that the potential is changing on time-scales longer than the time it takes for a characteristic star to traverse its orbital segment, our method could provide an accurate ``snapshot" of the present day potential. For a more sparsely populated stream, the measured orbital segments become larger, and our analysis becomes more sensitive to the evolving potential. In this regime, we might estimate a mean time-dependent potential, since as stars continue along their orbits the stream could deform. Indeed, the Milky Way halo may be evolving as a result of (e.g.) the LMC and its dark matter wake \citep{2019ApJ...884...51G,2021ApJ...919..109G,2021Natur.592..534C, 2022arXiv220501688L} or dark matter figure rotation predicted by $\Lambda \rm{CDM}$ \citep{1992ApJ...401..441D,2004ApJ...616...27B,2007MNRAS.380..657B,2021ApJ...910..150V}. These effects may lead to detectable morphological deviations in stellar streams, though further work is still needed to explore the impact of these time-dependent effects on the local dynamics of streams. In a future work, we intend to apply our method to simulated streams generated in a representative time-dependent potential.

Dark matter subhalos can create gaps in stellar streams as they orbit throughout the galaxy. The GD-1 stream in particular shows signs of such an encounter, due to a gap and spur of stars which bifurcate the stream \citep{2019ApJ...880...38B, deBoer2020wiggles}. Subhalo interactions which lead to gaps along a stream are not particularly challenging to handle in our analysis, and do not need to be modeled separately. Rather, we have demonstrated in \S\ref{sec: orbit_fitting} that our analysis can interpolate over missing segments of a given stream. Spur-like features that lead to distinct arms can be modeled as independent streams with their own track. 

Our main analysis has considered measurements of the full 6d phase-space coordinates of stream tracers.  This information is already available for several streams and will be measured with increasing precision across a larger number of systems with the advent of (e.g) \textit{Gaia} DR3. However, we have also implemented a version of our analysis which constrains the galactic acceleration field using incomplete phase-space measurements with missing dimensions. This makes our approach applicable to the large volume of {\it Gaia}-detected streams with missing line-of-sight velocities \citep{2021ApJ...914..123I}. We expand upon the case of missing phase-space dimensions in \S\ref{sec: future}.

\subsection{Future Directions}\label{sec: future}
In this paper we have established a new avenue for constraining the galactic potential using measurements of stellar streams. We intend to apply our method to streams with measured 6D phase-space dimensions in a future work. This will enable highly localized estimates of the galactic acceleration field, independent of any particular functional form for the gravitational potential.

A large number of streams have been measured with {\it Gaia} (see e.g., \citealt{2019ApJ...872..152I,2020MNRAS.492.1370B}), though most stream stars have 5D astrometric solutions (two angular coordinates, 2 proper motions, and a parallax angle). Even then, parallax measurements are not necessarily reliable for dim stars in the stellar halo. We have explored a modification to our analysis which is able to constrain components of the galactic acceleration field along a given stream in the absence of certain phase-space dimensions. In particular, the modeling introduced in \S\ref{sec: method} is based on vector differential equations (e.g., Eq.~\ref{eq: acceleration_eqn}). These vector expressions can be decomposed into components, allowing us to model streams with incomplete phase-space observations. For instance, if line-of-sight velocities are unavailable then Eq.~\ref{eq: acceleration_eqn} can be re-written in terms of 3D positions and 2D perpendicular velocity components in the plane of the sky. We can then infer the component of the galactic acceleration vector in the plane of the sky. Without distances we can infer angular accelerations, however, one still needs to estimate the heliocentric distance to the source in order to subtract the solar reflex motion. This can be done using the Galactic Parallax method introduced in \cite{2009MNRAS.399L.160E}. Alternatively, photometric distances can be estimated (see, e.g., \citealt{2018ApJ...862..114S}). In this way, our analysis is widely applicable to the large number of stellar streams which have 4D astrometric solutions from {\it Gaia} \citep{2021ApJ...914..123I}. Furthermore, the upcoming {\it Gaia} DR3 will include low-resolution spectra for millions of stars, which will be utilized to estimate isochronal distances. These measurements will make our analysis applicable to a large number of streams, providing small-scale constraints on the shape and mass of the Milky Way dark halo. For simplicity, we do not analyze mock-streams with incomplete phase-space observations in the present work. However, this is the subject of future work.

Observations of external galaxies have revealed tidal debris, thought to be remnants of accreted satellite galaxies (see, e.g., \citealt{2018ApJ...866..103K}). For M31 in particular, there is mounting evidence for accreted substructure including the Giant stellar stream and a shelf-feature also indicative of tidal disruption \citep{2001Natur.412...49I,2016ASSL..420..191F,2016MNRAS.458.3282C}. Radial velocities have been measured for a number of stars in the giant stellar stream, and photometric measurements have constrained a distance gradient along the stream \citep{2014ApJ...780..128I,2016MNRAS.458.3282C}. Our analysis can be applied to these measurements, to obtain a direct estimate of the line-of-sight accelerations along the Giant stellar stream in M31. The {\it Nancy Grace Roman Space Telescope} \citep{2015arXiv150303757S} is also expected to detect a large number of streams within M31 and other galaxies in the local volume \citep{2019ApJ...883...87P,2022ApJ...926..166P}. If radial velocities are obtained for these systems and a distance gradient can be estimated along the stream (using e.g. photometry; \citealt{2016MNRAS.458.3282C}), our method will provide an exciting avenue to measure the small-scale structure of the dark halo in external galaxies, placing the Milky Way in a broader context.

The methods developed in this paper are philosophically similar to recent work that discusses measuring the galactic acceleration field  using e.g. high-precision radial velocity measurements, pulsar timing arguments, or eclipsing binary star timing arguments \citep{2020ApJ...902L..28C,2021ApJ...907L..26C,2022ApJ...928L..17C}. In particular, we rely on changes in dynamical properties along a stream to estimate accelerations directly. Analogous to these works, our approach provides highly localized measurements of the galactic acceleration field in which the observed stream resides. Such localized measurements of the dark matter distribution in the stellar halo could prove useful for testing the standard cosmological model, $\Lambda\rm{CDM}$, on small-scales \citep{2015PNAS..11212249W} in a way that does not rely on generative prescriptions for stream formation, stringent energy requirements, and analytic user-defined models for the galactic potential. Comparing our analysis with other small-scale measurements of the dark matter distribution (e.g. gravitational lensing) will enable a test of systematic errors which may bias such highly localized constraints, since different methods are not necessarily subject to the same systematic errors. Our method therefore provides a potentially useful addition to techniques that measure galactic accelerations directly. 

In \S\ref{sec: limitations} we have discussed the limitations of our analysis in a time-dependent potential. Time dependence can lead to a significant misalignment between the local morphology of a stream and the trajectories of its tracers \citep{2019MNRAS.487.2685E}. This poses a potential challenge for our analysis. We have begun a preliminary exploration of a modification to our approach that can estimate accelerations even in the presence of such misalignments, by allowing for the possibility that the stream track is not directly coupled to stellar velocities. As discussed in \S\ref{sec: limitations}, we do anticipate that a large number of streams are strong candidates for the analysis presented in this paper already, so we intend to explore further generalizations in a future work.

The aim of this work is to introduce and demonstrate a new method for constraining galactic accelerations from measurements of stellar streams. As such, we have not incorporated a detailed discussion of uncertainty quantification using our method. From a preliminary analysis, we have implemented variational inference \citep{2021arXiv210813083G,2015arXiv150602142G} as a technique to provide estimates of model uncertainty on the inferred acceleration field along a stream. Characterizing model uncertainty is especially important on real data, where the ground-truth potential is unknown and the measurements themselves are uncertain. We therefore postpone detailed modeling of uncertainties to a future work, though we note that variational methods make uncertainty quantification using neural networks feasible. We have also already laid the groundwork to incorporate measurement errors in \S\ref{sec: fitting_streams}.

Finally, we conclude by considering a future application of our approach to systems beyond stellar streams. In particular, our method for estimating galactic accelerations requires the existence of a coherent structure in phase-space, with the property that, on average, nearby tracers have similar energies at slightly different phase-angles. In the galactic disk, level-sets of constant element abundances have been shown to roughly delineate stellar orbits \citep{2021ApJ...910...17P}. This is expected for a well-mixed equilibrium population, where abundances are analogous to orbital actions and an isoabundance contour is distributed over a range of conjugate angles. Indeed, this formulation has been applied to estimate the mass of the Milky Way disk in \cite{2021ApJ...910...17P}. In addition to this measurement, if an isoabundance curve can be estimated in phase-space using our approach, it would provide a new avenue to constrain the galactic acceleration field in the Milky Way disk without imposing functional priors on the form of the disk potential. We intend to explore this exciting application in a future work. 

\section*{Acknowledgements}
We are grateful to David Spergel, Kathryn Johnston, Adrian Price-Whelan, Julianne Dalcanton, Ana Bonaca, Robyn Sanderson, Shaunak Modak, Lachlan Lancaster, Camryn Phillips, Maureen Iplenski and the members of the Cambridge Streams Club for illuminating discussions that helped to improve this work. We also thank the anonymous referee for their valuable feedback on the manuscript.

\software{Numpy \citep{ harris2020array}, Matplotlib \citep{Hunter:2007}, PyTorch \citep{NEURIPS2019_9015}, SciPy \citep{2020SciPy-NMeth}, Gala \citep{gala,adrian_price_whelan_2020_4159870}, Astropy \citep{astropy:2018}, NEMO \citep{1995ASPC...77..398T}. }

\appendix
 \begin{figure*}[tp!]
    \centering
    \includegraphics[scale=0.6]{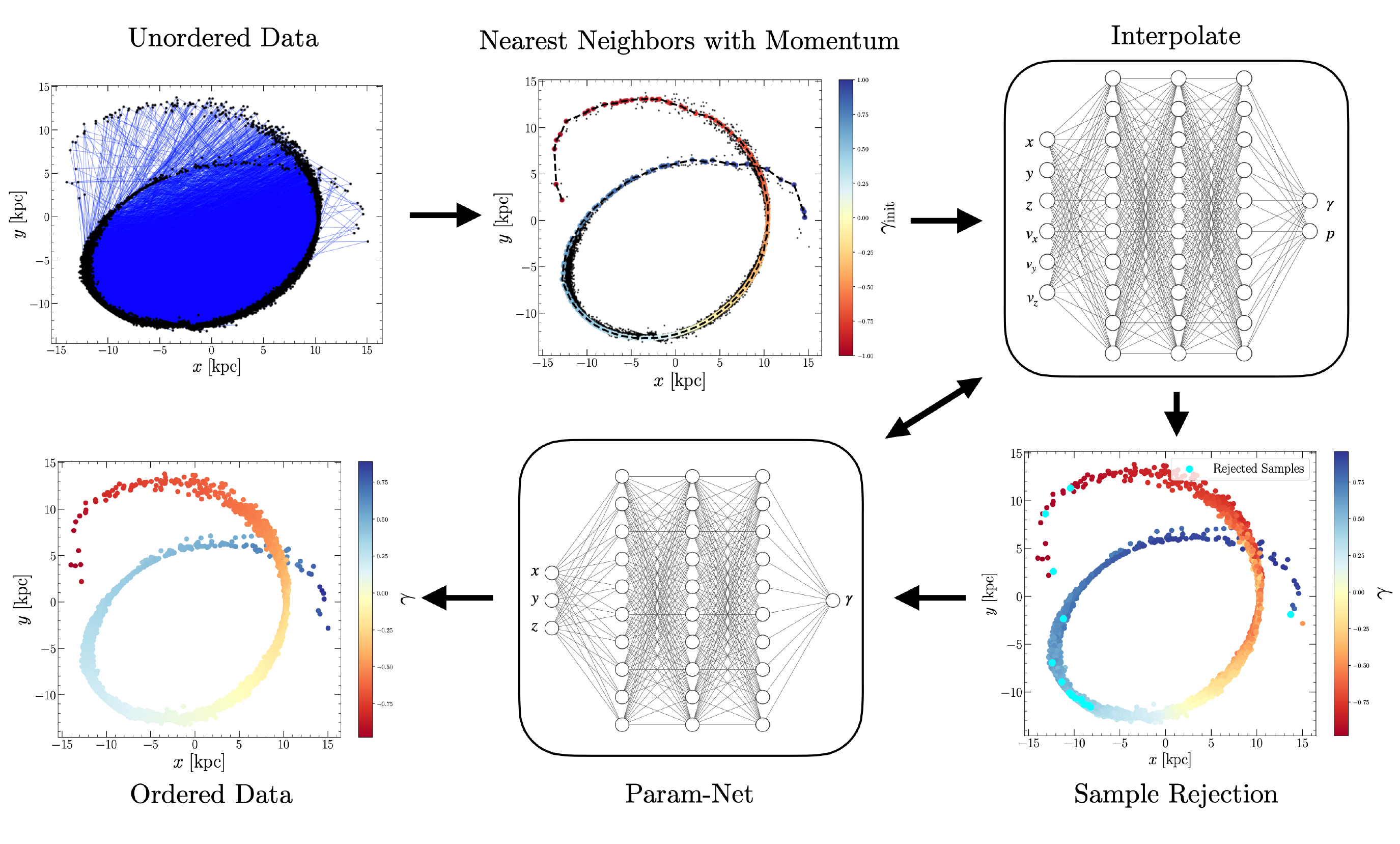}
    \caption{Schematic outline of stream unwrapping procedure. {\it Top Left:} Input phase-space data, where subsequent points in the dataset are connected by blue lines. {\it Top Middle:} We implement a nearest neighbors algorithm to connect tracers which have a small euclidean distance in phase-space. There is an added momentum condition,  which encourages the nearest neighbor graph edges in position space to fall roughly tangent to the local direction of motion of the stream. The stream is color-coded by the initial ordering, $\gamma_{\rm init}$. {\it Top Right:} The momentum condition inevitably leaves out a number of stream members from the nearest neighbors graph, requiring a flexible interpolation to assign an ordering (i.e., $\gamma$-value) to tracers which were initially left out of the graph. Interpolation is performed with a neural network, which outputs a membership probability for tracers in addition to the ordering parameter, $\gamma$. {\it Bottom Right:} Output of the interpolation neural network, color-coded by the interpolated $\gamma$-values across the stream. Blue points indicate tracers with the lowest probability membership, though all stars have membership probability $>0.99$ since they are stream members. {\it Bottom Middle:} The interpolation neural network (top right) is connected to a new neural network called Param-Net, which takes the ordering parameter $\gamma$ as an input, and outputs the mean position along the stream. $\gamma$-values are further refined using the same momentum condition as before: namely, that increasing $\gamma$ corresponds to motion along the stream from trailing to leading arm. {\it Bottom Left:} Example output of the final unwrapped stream, color-coded by the final $\gamma$-values.   }\label{fig: Param_Net}
\end{figure*}

\section{Unwrapping Stellar Streams: Determining $\gamma$}\label{App: Unwrap}
The stream track described in \S\ref{sec: fitting_streams} is a function of the scalar parameter $\gamma$, which encodes position along the track. For streams which do not wrap onto themselves, an on-sky position angle can be used to assign an ordering to stream stars. We denote the ordered set of $N$ stream tracers with $\{T_1, T_2, ..., T_N\}$, where $T_i$ represents the $i^{\rm th}$ tracer. We choose the convention that increasing $i$ corresponds to motion along the trajectory of the stream's orbit in phase-space (rather than against this trajectory). Once an ordering has been determined, a scalar parameter $\gamma_i$ can be assigned to each tracer, such that 
\begin{equation}\label{eq: ordered_set}
\begin{split}
     \mathrm{for \ the \ ordered \ set} \ &\{ T_1, T_2,..., T_N\},\\
    &\gamma_1 < \gamma_2 < ... < \gamma_N.
\end{split}
\end{equation}
In this way, $\gamma$ is analogous to a phase-angle along the ensemble of orbits that streams delineate. Because only the rank-ordering of $\gamma_i$ is meaningful, the numerical value is ambiguous. We therefore adopt the simple convention $\gamma_i \in [-1,1]$. In practice, we divide this interval by the number of stream tracers, such that $\gamma_{i+1} - \gamma_{i}$ is a positive constant.

Eq.~\ref{eq: ordered_set} implies that once the ordering $\{T_1,...,T_N\}$ is fixed, it is straightforward to assign a $\gamma_i$ to each tracer. Determining a suitable ordering, however, is more challenging. To determine an ordering for stream tracers and thus their corresponding $\gamma$-values, we adopt a multi-step process illustrated in Fig.~\ref{fig: Param_Net}. We describe the panels of this figure sequentially to highlight our approach.

\subsection{Nearest Neighbors with Momentum}
An unordered set of phase-space measurements is illustrated in the top left panel of Fig.~\ref{fig: Param_Net}, where subsequent tracers in the dataset are connected by a blue line. The result is a graph with tracers as nodes, though subsequent connections are randomized. We assume that all streams in our analysis are initialized in this way (i.e. with random ordering). For this particular stream a suitable ordering cannot be prescribed using an on-sky position angle alone, since the stream wraps onto itself. This differs from the stream in Fig.~\ref{fig: Stream_Sample}, which can be parametrized in terms of $\phi_1$. Instead, we adopt a modified nearest neighbors algorithm which connects subsequent tracers in phase-space, with the momentum condition that graph-edges which connect neighboring particles roughly align with the local direction of motion of stream tracers in position space. We provide a pseudocode version of the algorithm in Algorithm~\ref{Algorithm}, where the momentum condition is enforced through the \texttt{CosSimDist} penalty term. We find that this term is redundant for simple stream morphologies. However, for streams which loop back onto themselves one or more times, the momentum condition helps break degeneracies in the nearest neighbors search. By adding the \texttt{CosSimDist} term in Algorithm~\ref{Algorithm}, nearby tracers with velocity vectors pointing along the local graph edge will be preferred; this helps ensure that the ordering induced by the algorithm does not jump from one stream arm to another at an intersection. Additionally, the momentum condition helps ensure that the graph will not be dominated by a few anomalous stars with atypical velocities.

We illustrate the result of the nearest neighbors with momentum algorithm in the top-middle panel of Fig.~\ref{fig: Param_Net}. This panel contains the same stream as in the top left panel. Now that an ordering has been induced for a number of particles which populate the graph, we are able to assign a $\gamma$ value to these tracers (Eq.~\ref{eq: ordered_set}). We denote these starting values with $\gamma_{\rm init}$, since they will momentarily be updated and smoothed over. The leading arm is illustrated in the blue-range of the color-bar, while the trailing arm is in the red-range. The dashed-line connects neighboring graph nodes, similar to the blue lines in the top left panel, though now ordered. From this panel, it can be seen that a number of tracer particles were left out of the graph. This is due to the momentum condition, which keeps the nearest neighbors algorithm moving forward along the local trajectory of stream tracers. In order to assign a $\gamma$-value to the remaining tracers which were left out of the graph, we utilize a neural network to perform a flexible interpolation. 

\begin{algorithm}
\SetKwInOut{Output}{Output}
\SetKwInOut{Parameters}{Parameters}
\DontPrintSemicolon
\KwData{Phase-space observations, \texttt{PhaseSpace}; Index of phase-space initialization coordinate, \texttt{IdxStart}; Initial phase-space index, corresponding to $(\boldsymbol{x}_0, \boldsymbol{v}_0)$; Tracer position vector array, $\boldsymbol{X}$; Set of phase-space coordinates at which the algorithm will terminate, \texttt{PhaseSpaceTerminate}.}
\Parameters{
Maximum number of iterations, $N_{\rm run}$; Maximum allowable phase-space distance between neighboring tracers, \texttt{maxDist}; Weight for momentum condition, $\lambda$.}
\Output{Ordered phase-space coordinates,  \texttt{PhaseSpaceOrdered}}
\Begin{
  \texttt{PhaseSpaceCurr} $\longleftarrow$  \texttt{PhaseSpace}[\texttt{IdxStart}]\;
  \texttt{xCurr}, \texttt{vCurr} $\longleftarrow \boldsymbol{x}_0, \boldsymbol{v}_0$\;
  \texttt{PhaseSpaceModify} $\longleftarrow \texttt{PhaseSpace}$.\texttt{remove}(\texttt{index} = \texttt{IdxStart})\;
  \texttt{PhaseSpaceOrdered} $\longleftarrow$ [ ]\;
  \texttt{PhaseSpaceOrdered}.\texttt{append}(\texttt{PhaseSpaceCurr})\;
  \While{i $< N_{\rm run}$}{
  \If{\rm{\texttt{PhaseSpaceCurr} \textbf{in} \texttt{PhaseSpaceTerminate}}}{\textbf{break}}\;
  \texttt{d0} $\xleftarrow[]{} \Vert\mathrm{\texttt{PhaseSpaceCurr} - \texttt{PhaseSpaceModify}} \Vert$\;
  \If{\rm{\texttt{d0.min}()} $>$ \rm{\texttt{maxDist}}}{\textbf{break}}\;
  \rm{\texttt{UnitDeltaPos}} $\longleftarrow \left(\boldsymbol{X} - \rm{\texttt{xCurr}}\right)/\Vert \boldsymbol{X} - \rm{\texttt{xCurr}} \Vert$\;
  \texttt{CosSimDist} $\longleftarrow$ 1 - \texttt{sum}$(\rm{\texttt{UnitDeltaPos}}*\frac{\rm{\texttt{vCurr}}}{\Vert \rm{\texttt{vCurr}}\Vert})$\;
  \texttt{d} $\longleftarrow $ \texttt{d0} + $\lambda$  \texttt{CosSimDist}\;
  \texttt{ind} $\longleftarrow$ \texttt{argmin}(\texttt{d})\;
  \texttt{PhaseSpaceOrdered}.\texttt{append}(\texttt{PhaseSpaceModify}[\texttt{ind}])\;
  \texttt{PhaseSpaceModify} $\longleftarrow$\texttt{PhaseSpaceModify}.\texttt{remove}(\texttt{index} =   \texttt{ind})\;
\texttt{PhaseSpaceCurr} $\longleftarrow$ \texttt{PhaseSpaceOrdered}[-1]
  } }
\caption{Nearest Neighbors with Momentum \label{Algorithm}}
\end{algorithm}

\subsection{Autoencoder Neural Network}\label{app: autoencoder}
From the ordered, incomplete set of stream tracers, we now discuss an interpolation method to assign a $\gamma$-value to stars initially left out of the graph. To determine a $\gamma$-value for these tracers, we utilize a neural network with an autoencoder architecture, where the bottleneck of the autoencoder is the $\gamma$-parameter.

At a conceptual level, the autoencoder is split into two parts. The first we call the Interpolation network (Fig.~\ref{fig: Param_Net}; top right panel), while the second we call Param-Net (Fig.~\ref{fig: Param_Net}; bottom row, middle panel). The interpolation network takes 6D phase-space coordinates as an input, and outputs a $\gamma$-value and a new scalar $p \in [0,1]$, which indicates stream membership probability. While including velocity components in the interpolation neural network is not strictly required, we find that it helps break degeneracies when unwrapping streams which cross through the same spatial position several times.

We adopt a two-step training process for fitting the parameters of the autoencoder. First, we train the interpolation neural network as follows. From the nearest neighbors with momentum algorithm, we have already determined an initial set of $\gamma$-values for a number of tracers; namely, $\gamma_{\rm init}$. We adopt these values as labels, and utilize the interpolation network to predict new labels for tracers which were left out of the graph. For tracers which were connected in the initial graph, we assign them each a $p_i \equiv 1$, and train the neural network on these labels as well. We randomly sample points in a phase-space box which encloses the observed stream, assigning these randomized points $p \equiv 0$. The loss function---dependent on the neural network parameters, $\theta$---is constructed as follows,\begin{linenomath}
\begin{multline}\label{eq: loss_interpolate}
    \ell_{\theta}(\theta) \equiv \frac{1}{N}\sum_{i=1}^N\Bigg\{ \left(\gamma_{\theta}(\boldsymbol{x}_i,\boldsymbol{v}_i) - \gamma_{i,{\rm init}} \right)^2  \\ + \left(p_{\theta}(\boldsymbol{x}_i,\boldsymbol{v}_i) - 1 \right)^2 + \left(p_{\theta}(\boldsymbol{x}_{i,{\rm rand}},\boldsymbol{v}_{i,{\rm rand}})\right)^2 \Bigg\},
\end{multline}\end{linenomath}
where $(\boldsymbol{x}_i,\boldsymbol{v}_i)$ are the phase-space coordinates of tracers which populate the graph from the nearest neighbors algorithm, and $(\boldsymbol{x}_{i,{\rm rand}},\boldsymbol{v}_{i,{\rm rand}})$ are randomly drawn coordinates within a 6D phase-space box that encloses the stream. Training through this loss-function provides a mapping from phase-space to the scalar parameter $\gamma$. This enables us to assign a $\gamma$-value to tracers that were initially left out of the graph. Furthermore, the probability membership, $p_{\theta}$, is trained such that $p_{\theta} \approx 1$ will typically correspond to phase-space coordinates which are similar to those used in the training set (e.g. from the nearest neighbors analysis). Alternatively, $p_{\theta} < 1$ indicates that a particular tracer may not be assigned an accurate $\gamma$ at high-confidence, since it is dissimilar (i.e. out of distribution) compared to the training set. We achieve this behavior from the random sampling of $(\boldsymbol{x}_{i,{\rm rand}},\boldsymbol{v}_{i,{\rm rand}})$ in Eq.~\ref{eq: loss_interpolate}. Because stream members will typically occupy a narrow region in 6D phase-space, randomly sampled coordinates within a phase-space box are unlikely to frequently fall directly along the stream simultaneously in all 6 phase-space dimensions for a finite number of random samples. Additionally, stream members are already trained with $p_i \equiv 1$ over a narrow phase-space volume, thereby reducing contamination from the random sampling scheme. 

An example output of the interpolation neural network is illustrated in the bottom right panel of Fig.~\ref{fig: Param_Net}, where we have assigned a $\gamma$-value to all stream tracers using the interpolation, even for those which were originally left out of the nearest neighbors graph. All stream tracers are found to have $p > .99$, indicating high probability membership. We plot the top $\sim 20$ tracers with the lowest probability membership in the cyan points. Stars in the extended tidal tails have the lowest probability membership, though stars more centrally located along the stream have similar, higher membership probabilities as expected. 

We next consider the second half of the autoencoder, Param-Net (Fig~\ref{fig: Param_Net}; bottom row, middle panel). Param-Net characterizes the mapping from $\gamma$ to position space, $(x,y,z)$. Param-net provides the initial seed for the stream track neural network discussed in \S\ref{sec: fitting_streams}. The stream track neural network differs from Param-Net in that its $\gamma$-values are treated as fixed. We again utilize a fully-connected multi-layer perceptron neural network for Param-Net. If we treat the $\gamma$-values assigned by the Interpolation neural network in the top right panel of Fig.~\ref{fig: Param_Net} as fixed values, learning the mapping from $\gamma \xrightarrow[]{} \boldsymbol{x}$ is straightforward, since each stream tracer already has its own $\boldsymbol{x}_i$ and estimated $\gamma_i$ from the Interpolation network. However, a careful inspection of the particles in the bottom-right panel of Fig.~\ref{fig: Param_Net} reveals that some tracers have been misordered. For instance, there is a lone orange tracer at $(x,y)\approx (15,-4)$ which should have been assigned a $\gamma$ value with the opposite sign (e.g. $\gamma \approx 1$). This is an artifact of our incomplete training set for $\gamma_{\rm init}$, which is constructed from the nearest neighbors search and inevitably leaves a population of tracers disconnected from the graph.

We address these misclassifications by reinforcing the same momentum condition used in the nearest neighbors search, though in the context of neural network training. That is, we smooth out the ordering of stream tracers inferred by the Interpolation network by encouraging the momentum condition that $d\boldsymbol{x}(\gamma_i)/d\gamma$ is parallel to the direction of motion of tracers, $\boldsymbol{v}_i/\Vert \boldsymbol{v}_i \Vert$.
This is implemented by connecting the Interpolation network to Param-Net through the shared $\gamma$-parameter (Fig.~\ref{fig: Param_Net}; diagonal arrow), reducing two independent neural networks down to one through an autoencoder structure. $\gamma$-values are fine tuned by the simultaneous training of these neural networks, through the loss function\begin{linenomath}
\begin{multline}\label{eq: loss_auto_encoder}
    \ell_{\theta}\left(\theta\right) \equiv   \sum_{n=1}^N \Bigg[ \Vert \boldsymbol{x}_n - \boldsymbol{x}_{\theta}\left(\gamma_{\theta}(\boldsymbol{x}_n,\boldsymbol{v}_n)\right)  \Vert^2  \\ + \lambda \Vert \boldsymbol{T}_n - \boldsymbol{T}_{\theta}\left((\gamma_{\theta}(\boldsymbol{x}_n,\boldsymbol{v}_n)\right) \Vert^2 \Bigg],
\end{multline}\end{linenomath}
where $\boldsymbol{T}_n$ and $\boldsymbol{T}_\theta$ are tangent vectors, defined in \S\ref{sec: fitting_streams}, and $\lambda$ is a hyper-parameter which we set to $\sim 10^2$. Adjusting the $\gamma$ values inferred by the Interpolation neural network through this loss function, which simultaneously tunes the parameters of Param-Net, leads to a more accurate, smooth ordering of stream tracers illustrated in the bottom left panel of Fig.~\ref{fig: Param_Net}.  In this panel, the misplaced tracer at $(x,y)\approx(15,-4)$ in the bottom right panel has been assigned a reasonable $\gamma$-value. 

Once a smooth ordering of tracers has been achieved (bottom left panel of Fig.~\ref{fig: Param_Net}), $\gamma$-values are fixed in the analyis and no further adjustments to the ordering of stream tracers are made. The stream track and track speed neural networks are trained on these fixed $\gamma$-values. 

\section{Neural Network Architectures}\label{App: Architectures}
In this section we provide a description of the neural network architectures utilized in this work. 

\subsection{Stream Track}
The stream track neural network is a fully connected multilayer perceptron  with a single input node ($\gamma$) and a 3-dimensional output representing the cartesian $(x,y,z)$ coordinates of the stream track. We use 3 hidden layers consisting of 100 nodes each, with $\tanh$ activation functions. The output node is a linear combination of weights and biases, and is not mapped through a non-linear activation function. 

Rather than training on the raw $(x,y,z)$ coordinates of stream stars, we standardize each dimension of the training data by its mean $\mu$ and standard deviation $\sigma$. That is, for each spatial dimension $x_i$ we scale the training data as follows:
\begin{equation}\label{eq: scale_network}
    x_i \longrightarrow \frac{x_i - \mu_i}{\sigma_i} \equiv x_i^{\rm scaled}.
\end{equation}
We apply the inverse transformation of Eq.~\ref{eq: scale_network} (e.g. $x_i^{\rm scaled}\sigma_i + \mu_i$) to infer the unscaled stream track. This neural network has an identical structure to Param-Net, previously described in Appendix~\ref{App: Unwrap}.

\subsection{Track Speed}
The track-speed neural network is a fully connected multilayer perceptron with a single input node ($\gamma$) and a scalar output node representing the average local speed of stream stars, $v \equiv \Vert \boldsymbol{v}\Vert$. We use 3 hidden layers consisting of 36 nodes each, with $\tanh$ activation functions. The output node is a linear combination of weights and biases, and is mapped through an exponential to yield a positive estimate of the mean speed. 

\subsection{Interpolation Neural Network}
The interpolation neural network is utilized to estimate the parameter $\gamma$ as a function of position $\boldsymbol{x}$. It is described in Appendix~\ref{app: autoencoder}. The interpolation network is a fully connected multilayer perceptron with 6 input nodes representing the phase-space coordinates $(\boldsymbol{x},\boldsymbol{v})$, and two output nodes representing the $\gamma$-parameter and a stream membership probability, $p$. We use 3 hidden layers consisting of 100 nodes each, with $\tanh$ activation functions. We adopt the convention that $\gamma \in [-1,1]$, by applying a $\tanh$ activation function to the output node corresponding to the $\gamma$-parameter. For the probability parameter $p \in [0,1]$, we use a sigmoid activation function. 

\vspace{-.1cm}
\subsection{Potential Neural Network}
The neural network used to estimate the gravitational potential in \S\ref{sec: analytic_potential} is a fully connected multilayer perceptron with 3 input nodes representing position $\boldsymbol{x} \in \mathbb{R}^3$ relative to the galactic center, and one output node representing the gravitational potential. We use 3 hidden layers, consisting of 36 nodes each with $\rm{tanh}$ activation functions. The output node is also mapped through a $\rm{tanh}$ activation function, which is then multiplied by a positive amplitude parameter $A$. The amplitude is also tuned during neural network training. One subtlety of inferring the gravitational potential from accelerations is that the potential $\Phi$ and $\Phi + \rm{const.}$ are both equally valid. We therefore impose the condition that at the origin, $\boldsymbol{x}=\boldsymbol{0}$, the potential is zero. We implement this by passing the point $\boldsymbol{x} = \boldsymbol{0}$ through the potential neural network for each evaluation point, multiplying by the amplitude $A$, and subtracting the resulting value from the neural network output corresponding to the arbitrary position $\boldsymbol{x}$. This ensures that at the origin, the estimated potential will always be zero since all outputs are relative to the potential at the origin. In practice, using a $\rm{tanh}$ activation function scaled by amplitude $A$ for the output node is found to yield faster convergence to the estimated accelerations and more stable training, likely because mapping the output node through the function $A \tanh{(\cdot)}$ reduces the need for particularly large weights and biases to reproduce the estimated accelerations. Similar results can be achieved without mapping the output node through the $\rm{\tanh{}}$ function, though fractional errors are found to be slightly higher after the same training period compared to the case with the $\rm{\tanh}$ output. 

\section{Existence of a Physical Stream Track}\label{app: physical_track}
In this work we estimate a one-dimensional track that characterizes the mean position and dynamics of a stream. The output of our analysis is an acceleration vector sampled along the stream track, namely, $\boldsymbol{a}(\gamma)$. From this acceleration vector we can compute the derivative $d\boldsymbol{a}/d\gamma$. However, because we infer what is effectively a ``slice" through the full acceleration field, calculating derivatives like $\partial\boldsymbol{a}/\partial x_i$ in general (for cartesian coordinate $x_i$) is not possible. It is therefore interesting to consider under what conditions the inferred accelerations are physical. In particular, for a positive mass density we expect $\nabla \cdot \boldsymbol{a} \leq 0$. For a conservative gravitational force, we also expect $\nabla \times \boldsymbol{a} = \boldsymbol{0}$. Indeed, $d\boldsymbol{a}/d\gamma$ (which we can compute directly) is related to the derivatives involved in computing the divergence and cross product above. Namely,
\begin{equation}\label{eq: acc_components_derivs}
    \frac{da_i}{d\gamma} = \sum_j \frac{da_i}{dx_j}\frac{dx_j}{d\gamma} = \nabla a_i \cdot \frac{d\boldsymbol{x}}{d\gamma}
\end{equation}
for the $i^{\rm th}$ cartesian component of the acceleration $\boldsymbol{a} = (a_1,a_2,a_3)$. 
We define the unit vector 
\begin{equation}\label{eq: unit_tang}
    \boldsymbol{T} \equiv \frac{d\boldsymbol{x}/d\gamma}{\Vert d\boldsymbol{x}/d\gamma\Vert},
\end{equation}
which is tangent to the stream track, $\boldsymbol{x}(\gamma)$. Dividing Eq.~\ref{eq: acc_components_derivs} by $\Vert d\boldsymbol{x}/d\gamma\Vert$, one obtains
\begin{equation}\label{eq: directional_deriv}
    \frac{da_i}{d\gamma} \Big/ \Big\Vert\frac{d\boldsymbol{x}}{d\gamma}\Big\Vert = \nabla a_i \cdot \boldsymbol{T}(\gamma). 
\end{equation}
Because the left-hand-side of Eq.~\ref{eq: directional_deriv} can be calculated explicitly using our method, we have access to the directional derivative of each acceleration component along the stream track. 
We now consider working in an orthogonal coordinate system for which one axis is parallel to the stream track for each evaluation point $\gamma$. A common choice is the Frenet–Serret frame or TNB frame \citep{stoker_1969}, which is defined by the tangent unit vector $(\boldsymbol{T}$; defined in Eq.~\ref{eq: unit_tang}), normal unit vector ($\boldsymbol{N} \equiv \boldsymbol{T}^\prime(\gamma)/\Vert\boldsymbol{T}^\prime(\gamma)\Vert$), and binormal unit vector ($\boldsymbol{B} \equiv \boldsymbol{T} \times \boldsymbol{N})$ along a parametrized curve in 3-dimensions. Collectively, these unit vectors construct an orthogonal right-handed coordinate system. In our context, each unit vector is a function of the scalar parameter $\gamma$, such that the orientation of the coordinate system changes along the stream. For the remainder of this section, we assume that $\nabla$ is the usual gradient operator though expressed in the TNB frame. That is, $\nabla \equiv \boldsymbol{T}\frac{\partial }{\partial T} + \boldsymbol{N}\frac{ \partial}{\partial N} + \boldsymbol{B}\frac{\partial}{\partial B} $. In this frame, Eq.~\ref{eq: directional_deriv} reduces to
\begin{equation}
    \nabla a_i \cdot \boldsymbol{T} = \left( \boldsymbol{T}\frac{\partial a_i}{\partial T} + \boldsymbol{N}\frac{\partial a_i}{\partial N} + \boldsymbol{B}\frac{\partial a_i}{\partial B}  \right) \cdot \boldsymbol{T} = \frac{\partial a_i}{\partial T},
\end{equation}
where $i$ can be $T, N$ or $B$. We therefore cannot compute derivatives with respect to any quantity other than $T$. This means that for a given stream track, it will always be possible to construct an acceleration field that satisfies
\begin{equation}\label{eq: TNB_div}
    \nabla \cdot \boldsymbol{a} = \frac{\partial {a}_T}{\partial T} + \frac{\partial {a}_N}{\partial N} + \frac{\partial {a}_B}{\partial B}  \leq 0.
\end{equation}
This is because we only constrain the term in the divergence containing the variable $T$. The other two derivatives are not constrained by our analysis, and can always be chosen so as to satisfy Eq.~\ref{eq: TNB_div}.

For the curl of $\boldsymbol{a}$, in the TNB frame we have\begin{linenomath}
\begin{multline}
    \nabla \times \boldsymbol{a} = \boldsymbol{T}\left(\frac{\partial a_B}{\partial N} - \frac{\partial a_N}{\partial B} \right) + \boldsymbol{N}\left(\frac{\partial a_T}{\partial B} - \frac{\partial a_B}{\partial T} \right)
    \\+ \boldsymbol{B}\left(\frac{\partial a_N}{\partial T} - \frac{\partial a_T}{\partial N} \right).
\end{multline}\end{linenomath}
Because we only constrain derivatives of the acceleration components with respect to $T$, one is guaranteed that the remaining derivatives can be chosen to satisfy $\nabla \times \boldsymbol{a} = \boldsymbol{0}$, provided that the stream track does not form a closed loop (for a closed loop track, Stokes' theorem can be used to enforce the curl-free condition. However, no such stream is expected to exist). We note that from the curl-free assumption, we can compute the derivatives $\partial a_T/\partial B$ and $\partial a_N/\partial T$. However, these derivatives are not necessary in order to compute Eq.~\ref{eq: TNB_div}.

We emphasize that when we reconstruct the gravitational potential (\S\ref{sec: potential}) from the estimated accelerations $\boldsymbol{a}(\gamma)$ for several streams, we are able to impose a positive mass density as a physically motivated prior. That is, we impose $\nabla^2\Phi(\boldsymbol{x}) \geq 0$. We are able to calculate such terms since we model the full scalar potential, rather than a curve parameterized through it.

\clearpage
\bibliography{thebib}
\end{document}